\documentclass[final,3p,times]{elsarticle}
\usepackage{amssymb}
\usepackage[hidelinks, urlcolor=black, colorlinks=true, linkcolor=black, citecolor=black]{hyperref}
\usepackage{url}
\usepackage{bbding}
\usepackage{siunitx}
\usepackage{pifont}
\usepackage[table]{xcolor}
\usepackage{multirow,array}
\usepackage{makecell}
\usepackage{booktabs}

\usepackage{subcaption}
\usepackage[numbers]{natbib}
\usepackage[font=small]{caption}
\usepackage{amsmath}
\usepackage{acronym}\acrodef{3GPP}[3GPP]{3rd generation partnership project}
\acrodef{AI}[AI]{artificial intelligence}
\acrodef{IP}[IP]{Internet Protocol}
\acrodef{B5G}[B5G]{beyond 5G}
\acrodef{KPI}[KPI]{key performance indicator}
\acrodef{5G}[5G]{5th generation}
\acrodef{SLA}[SLA]{service-level agreement}
\acrodef{MAPE}[MAPE]{Mean Absolute Percentage Error}
\acrodef{5G/6G}[5G/6G]{5th and 6th generation}
\acrodef{QoS}[QoS]{quality of service}
\acrodef{OSPF}[OSPF]{Open Shortest Path Protocol}
\acrodef{AC}[AC]{admission control}
\acrodef{SDN}[SDN]{software-defined networking}
\acrodef{VM}[VM]{Virtual Machine}
\acrodef{VNF}[VNF]{Virtual Network Function}
\acrodef{PDB}[PDB]{packet delay budget}
\acrodef{API}[API]{application programming interface}
\acrodef{VTwin}[VTwin]{virtual twin}
\acrodef{GRU}[GRU]{gated recurrent unit}
\acrodef{NMSE}[NMSE]{normalized mean square error}
\acrodef{PTwin}[PTwin]{physical twin}
\acrodef{MPNN}[MPNN]{message passing neural network}
\acrodef{DT}[DT]{digital twin}
\acrodef{NDT}[NDT]{network digital twin}
\acrodef{KSWIN}[KSWIN]{Kolmogorov-Smirnov windowing}
\acrodef{GNN}[GNN]{graph neural network}
\acrodef{MLP}[MLP]{multilayer perceptron}
\acrodef{UDP}[UDP]{User Datagram Protocol}
\acrodef{D-ITG}[D-ITG]{Distributed Internet Traffic Generator}
\acrodef{ML}[ML]{machine learning}
\acrodef{HMPGNN}[HMPGNN]{heterogeneous message passing graph neural network}
\acrodef{LSTM}[LSTM]{Long Short-Term Memory}
\acrodef{DGAT}[DGAT]{dynamic graph attention network}
\acrodef{ONOS}[ONOS]{Open Network Operating System}
\acrodef{LCM}[LCM]{lifecycle management}
\acrodef{TCP}[TCP]{Transmission Control Protocol}
\acrodef{REST}[REST]{REpresentational State Transfer}
\acrodef{RAN}[RAN]{radio access network}
\acrodef{CPS}[CPS]{cyber-physical system}

\journal{Computer Networks}

\begin{document}

\begin{frontmatter}

%% Title, authors and addresses

%% use the tnoteref command within \title for footnotes;
%% use the tnotetext command for theassociated footnote;
%% use the fnref command within \author or \affiliation for footnotes;
%% use the fntext command for theassociated footnote;
%% use the corref command within \author for corresponding author footnotes;
%% use the cortext command for theassociated footnote;
%% use the ead command for the email address,
%% and the form \ead[url] for the home page:
%% \title{Title\tnoteref{label1}}
%% \tnotetext[label1]{}
%% \author{Name\corref{cor1}\fnref{label2}}
%% \ead{email address}
%% \ead[url]{home page}
%% \fntext[label2]{}
%% \cortext[cor1]{}
%% \affiliation{organization={},
%%             addressline={},
%%             city={},
%%             postcode={},
%%             state={},
%%             country={}}
%% \fntext[label3]{}

\title{Towards a Robust Transport Network
With Self-adaptive Network Digital
Twin}

%% use optional labels to link authors explicitly to addresses:
%% \author[label1,label2]{}
%% \affiliation[label1]{organization={},
%%             addressline={},
%%             city={},
%%             postcode={},
%%             state={},
%%             country={}}
%%
%% \affiliation[label2]{organization={},
%%             addressline={},
%%             city={},
%%             postcode={},
%%             state={},
%%             country={}}

\author[lasse]{Cláudio Modesto\corref{cor1}} %% Author name
\ead{claudio.barata@itec.ufpa.br}
\author[lasse]{João Borges}
%% Author name
\author[lasse]{Cleverson Nahum} %% Author name
\author[lasse]{Lucas Matni} %% Author name
\author[unisinos]{Cristiano Bonato Both} %% Author name
\author[ufg]{Kleber Cardoso} %% Author name
\author[lasse]{Glauco Gonçalves} %% Author name
\author[lasse]{Ilan Correa} %% Author name
\author[ericsson]{Silvia Lins} %% Author name
\author[ericsson]{Andrey Silva} %% Author name
\author[lasse]{Aldebaro Klautau} %% Author name

\cortext[cor1]{Correspoding author.}
%% Author affiliation
\affiliation[lasse]{organization={LASSE, Federal University of Pará},%Department and Organization
            addressline={Belém},
            postcode={66075-110},
            country={Brazil}}

\affiliation[ufg]{organization={Institute of Informatics, Federal University of Goiás},%Department and Organization
            addressline={Goiânia},
            postcode={74001-970},
            country={Brazil}}
            
\affiliation[unisinos]{organization={Applied Computing
Graduate Program, University of Vale do Rio dos Sinos},%Department and Organization
            addressline={São Leopoldo},
            postcode={93022-750},
            country={Brazil}}

\affiliation[ericsson]{organization={Ericsson Research},%Department and Organization
            state={São Paulo},
            addressline={Indaiatuba},
            postcode={13337-300},
            country={Brazil}}

%% Abstract
\begin{abstract}
The ability of the \ac{NDT} to remain aware of changes in its physical counterpart, known as the \ac{PTwin}, is a fundamental condition to enable timely synchronization, also referred to as \textit{twinning}. In this way, considering a transport network, a key requirement is to handle unexpected traffic variability and dynamically adapt to maintain optimal performance in the associated virtual model, known as the \acf{VTwin}. In this context, we propose a self-adaptive implementation of a novel \ac{NDT} architecture designed to provide accurate delay predictions, even under fluctuating traffic conditions. This architecture addresses an essential challenge, underexplored in the literature: improving the resilience of data-driven \ac{NDT} platforms against traffic variability and improving synchronization between the \ac{VTwin} and its physical counterpart. Therefore, the contributions of this article rely on \ac{NDT} lifecycle by focusing on the operational phase, where telemetry modules are used to monitor incoming traffic, and concept drift detection techniques guide retraining decisions aimed at updating and redeploying the \ac{VTwin} when necessary. We validate our architecture with a network management use case, across various emulated network topologies, and diverse traffic patterns to demonstrate its effectiveness in preserving acceptable performance and predicting \ac{QoS} metrics under unexpected traffic variation, such as delay and jitter. The results in all tested topologies, using the normalized mean square error as the evaluation metric, demonstrate that our proposed architecture, after a traffic concept drift, achieves a performance improvement in per-flow delay and jitter prediction of at least $64\%$ and $21\%$, respectively, compared to a configuration without \ac{NDT} synchronization.
\end{abstract}

%%Graphical abstract
%\begin{graphicalabstract}
%\vspace{4cm}
%\centering
%\includegraphics[scale=0.%55]{Figures/figure_1.pdf}
%\end{graphicalabstract}

%% Keywords
\begin{keyword}
%% keywords here, in the form: keyword \sep keyword
Concept drift, \acf{GNN}, \acf{LCM}, twinning.
%% PACS codes here, in the form: \PACS code \sep code

%% MSC codes here, in the form: \MSC code \sep code
%% or \MSC[2008] code \sep code (2000 is the default)

\end{keyword}

\end{frontmatter}

%% Add \usepackage{lineno} before \begin{document} and uncomment 
%% following line to enable line numbers
%% \linenumbers

%% main text
%%

%% Use \section commands to start a section
\section{Introduction} \label{sec:introduction}

\Acf{NDT} is a platform designed to aid in the management and automation of physical network communication tasks based on interactions with its virtual counterpart, also called \acf{VTwin}, at a certain periodicity and fidelity~\cite{barricelli2019, ohlen2022}. This relation allows the network operator to emulate different scenarios in network planning to perform a \textit{what-if} analysis~\cite{varela2023, abenathar2025}. This analysis aims to evaluate the impact of management changes in the network infrastructure when these tests are impossible due to logistical issues or costs that render them unfeasible, such as in production environments~\cite{Almasan2022DigitalTN, ohlen2022}. For that reason, this platform is considered an enabling technology to optimize \ac{5G/6G} network communication~\cite{ohlen2022}, providing improved solutions to problems such as slicing management~\cite{Wang2022, farreras2025gnnetslice}, routing optimization~\cite{rusek2020routenet, almasan2022}, resource power allocation~\cite{varela2023}, and admission control~\cite{messaoudi2023}.

Use of an \ac{NDT} platform in a production environment can create several challenges, particularly concerning the long-term accuracy of the \ac{DT}. It is essential to account for variability in incoming data, ensuring that the \ac{VTwin} continually adapts to reflect real-world changes. When we consider a \ac{VTwin} neural network model to achieve this reliability, such as a \acf{GNN}~\cite{ferriol2023routenet}, it is fundamental to have a well-defined \acf{LCM} that organizes all phases, including the \textit{planning}, \textit{training}, and \textit{operational} aspects of the \ac{NDT} platform~\cite{bariah2024interplay, Ahmadi2021, Kherbache2021}. When a deployed \ac{NDT} is used during the operational phase, an important research question arises concerning the synchronization between the \ac{VTwin} and the \acf{PTwin}. Synchronization issues are widely explored in the general context of \acp{DT}, considering applications at factory workstations, the industrial Internet of Things, power systems, etc.~\cite{Alghamdi2024, Hribernik2021}. However, these questions are rarely explored within the specific context of a transport network \ac{NDT} based on \ac{GNN}. Given that the transport network plays a crucial role as the interconnection between the core network and the \ac{RAN}~\cite{farreras2025gnnetslice, Morais2023}, and the extensive body of literature supporting data-driven approaches using \ac{GNN}~\cite{rusek2020routenet} for modeling and representing transport networks, it becomes possible to investigate particular challenges related to twin synchronization, such as the methods used to achieve it and the underlying synchronization patterns~\cite{Alghamdi2024}.

In this sense, we propose an enhanced architecture grounded in the core lifecycle principles of \ac{NDT} to address these challenges, focusing on maintenance and resilience. The architecture is designed to support a robust platform capable of adapting to traffic variability, also referred to as \textit{covariate drift}, a special case of concept drift~\cite{baier2023, Ameur2025}, which can arise from disruptions such as congestion, unreachable switches, broken links, or shifts in traffic patterns. Our approach emphasizes synchronization between the \ac{VTwin} and the \ac{PTwin} aimed to assist data-driven approaches remain up-to-date whenever a input distribution change is detected. We adopted a data-driven synchronization pattern \cite{Alghamdi2024} managed through a concept drift detector that monitors network traffic for changes. When significant deviations are detected, a retraining process is triggered to realign the \ac{VTwin} with the current state of the \ac{PTwin}. Since our simulations employ a supervised learning setup, where the ``ground truth'' total delay and jitter are known, we evaluate and validate the generalization performance of the \ac{VTwin} in the proposed architecture as the \ac{PTwin} experiences dynamic changes in traffic intensity and patterns. The results of the proposed experiments demonstrate a performance improvement in all tested emulated topologies with different network traffic patterns, using the \acf{NMSE} as the evaluation metric. This shows that our proposed architecture achieves an improvement in total per-flow delay and jitter prediction of at least $64\%$ and $21\%$, respectively, after a traffic concept drift. In some other topologies, this improvement is greater than $99\%$ and $98\%$, respectively. Finally, with these improvements, we propose a final evaluation in a use case application based on \ac{SLA} monitoring to demonstrate the importance of \ac{NDT} synchronization, considering high-level metrics for network planning, such as the number of \ac{SLA} violations.

Our main contributions to this article are:
\begin{itemize}
    \item A novel \ac{NDT} architecture incorporating \ac{LCM} principles during the operational phase, such as the re-deployment of an updated \ac{VTwin} model when necessary. This contribution aimed to assist data-driven approaches (used to predict per-flow delay and jitter) remain up-to-date whenever a input distribution change is detected.
    
    \item As part of our architecture, a concept drift detection module is integrated to monitor incoming network traffic features. This \ac{NDT} module enables the analysis of traffic behavior to make informed decisions about when to update the \ac{VTwin} model.
    
    %\item Relevant \ac{NDT} experiments that account for environments with varying traffic conditions, including changes in traffic patterns and levels of congestion.

    \item A use case application for network management related to \ac{SLA} violation prediction compares the performance of the proposed application with and without \ac{NDT} synchronization.
\end{itemize}

The remainder of this article is organized as follows: in Section \ref{sec:Related Work} we compare our work with different proposed data-driven \ac{NDT} in the literature, considering various levels of implementation in the transport network domain and their characteristics related to \ac{NDT} supported modules, such as the synchronization module. In Section \ref{sec: proposed NDT}, we present the proposed enhanced \ac{NDT} architecture, taking into account the main functional requirements of \ac{NDT}, and show the essential building blocks of each, such as the \ac{VTwin} architecture, the synchronization methods based on concept drift detection techniques, and complementary modules. In Section \ref{sec:experiments}, we detail the experiments conducted to evaluate the proposed methods for twin synchronization and present the key evaluation results across various topologies and traffic patterns, as well as a use case with \ac{SLA} monitoring. Finally, Section \ref{sec:conclusion} summarizes the main contributions of this work and outlines directions for future research. 

\newpage
\section{Related Work} 

\newcommand{\xmark}{\ding{55}}%
\newcommand{\cmark}{\ding{51}}%
\label{sec:Related Work}

Recent studies point \acp{DT} to an important technology for the \ac{5G/6G} networks~\cite{nguyen2021digital, khan2022digital, tao2024wireless, sheraz2024comprehensive, liu2024digital}. This aligns with their need to establish reliable environments for developing and testing their critical parts, such as wired environments~\cite{messaoudi2023}, where \acp{DT} can provide excellent support. However, during this wave of adoption, the intersection of \ac{DT} and networking—still an emerging area of study—led to a diverse range of approaches. Many of these did not align fully with the definitions of an \ac{NDT} as systematized by standards organizations~\cite{itu_digital_twin, ndt_ietf} and industry stakeholders~\cite{ohlen2022, ORANRR}.  
Consequently, related work found in the literature presents limited analyzes of practical aspects. For example, regarding the system lifecycle or the twinning rates, the latter refers to the rate of synchronization between the \ac{PTwin} current state and its corresponding representation in the \ac{VTwin}~\cite{jones2020}. More specifically, much of the literature regarding \acp{NDT} of the transport domain has focused on using \ac{ML}/surrogate model-based methods to generate the \ac{VTwin}. The increase in popularity of this \ac{DT} approach, which can also be called data-driven~\cite{bariah2024interplay}, can be traced back to the work of Rusek et al.~\cite{rusek2020routenet}, which established the usage of a \ac{GNN} model called RouteNet for predicting path-level metrics in routing optimization and network upgrade use cases. The usage of this model was further investigated by several works, which were built on the previously established RouteNet~\cite{happ2021graph, saravanan2022enabling, Farreras2023, li2023learnable, ferriol2023routenet, messaoudi2023, Modesto2024}. 

Recently, the work of Ferriol-Galmés et al.~\cite{ferriol2023routenet} utilized a modified version of RouteNet, called RouteNet-Fermi, to model network communication with various topologies, traffic distributions, and queuing configurations, enabling the prediction of packet loss, delay, and jitter. In this work, since all data is provided from a packet-level simulator, there are no two-way interactions between \ac{PTwin} and \ac{VTwin}. Therefore, the case presented in the article does not constitute a closed-loop \ac{NDT}. Moreover, the investigations did not address the possible performance degradation of the \ac{VTwin} in cases of data and/or concept drifts~\cite{lu2018conceptdrift}. In this context, the cases presented in this work and several others built on RouteNet~\cite{happ2021graph, saravanan2022enabling, Farreras2023, li2023learnable, ferriol2023routenet} did not address the robustness aspect of an \ac{NDT}, as they did not explore its lifecycle either.

The exception to this lack of closed-loop implementation in works based on RouteNet is the work of Messaoudi et al.~\cite{messaoudi2023}, Zalat et al.~\cite{Zalat2024}, and Aben-Athar et al.~\cite{abenathar2025}, which are, to the best of our knowledge, the first closed-loop environments considering physical layers that interact with an \ac{AI} model. The works in \cite{messaoudi2023, abenathar2025} feed an application module tailored to operate over the physical counterpart defined by virtual switches on Mininet~\cite{mininetsite}. The work in ~\cite{Zalat2024} took the same approach but considered a small-scale physical testbed. However, a noteworthy aspect of this work is that there are elements for twin synchronization that are based on a trigger-based online learning approach. The main difference between this work and our proposed architecture, apart from the use of online learning training, is that the authors specifically focused on exploiting the correlation between the predicted per-flow delay and the ground truth per-flow delay to update the states of the \ac{VTwin} based on those of the \ac{PTwin}. This correlation, when applied, can also be considered the simplest form to compare two twin states (outputs) without explicitly using a concept drift detector. However, a key drawback is the requirement for ground-truth labels to start a retraining process. For the other two works, despite closed-loop interactions between the physical and virtual domains, there is no \ac{DT} lifecycle management controlling the rate at which both parts synchronize and update their respective states, for instance, through a retraining process. In this context, the authors in~\cite{messaoudi2023, abenathar2025} ran the traffic only once, and after completing it, they performed no further operations to detect whether the \ac{VTwin} performance would remain acceptable.

In addition to these, other works in the literature either proposed purely GNN-based approaches, such as FlowDT \cite{ferriol2022flowdt} or GNNetSlice~\cite{farreras2025gnnetslice}, or adopted different \ac{ML} models as their \acp{VTwin}. For example, Lai et al.~\cite{lai2023deep} adopted a model named eConvLSTM; Shin et al.~\cite{shin2023network} used \ac{GNN} along with \ac{LSTM} and a data extraction approach~\cite{yu2023digital} that uses \ac{DGAT}. However, among these other works, only Shin et al.~\cite{shin2023network} and Yu et al.~\cite{yu2023digital} presented investigations on the \ac{NDT} \ac{LCM}, while none presented concept drift implementations.

Finally, in terms of the datasets used in these works, there is a noticeable concentration of studies \cite{Wang2022, saravanan2022enabling, farreras2025gnnetslice, shin2023network, Farreras2023} based on specific datasets derived from the OMNeT++ simulator \cite{Varga2001}. This trend presents advantages and limitations. On the one hand, using datasets from a common source provides a stable baseline, enabling fair comparisons across different approaches in terms of generalization performance. On the other hand, relying solely on pre-built datasets limits the exploration of more realistic aspects of \ac{NDT}, such as the aforementioned closed-loop operations or \ac{DT} synchronization, as there is no interaction with a real or emulated environment.

Given the current state of the literature on twin synchronization in the transport network domain, Table~\ref{tabRelatedWork} summarizes the works discussed in this section based on four primary characteristics: (i) closed-loop operation, referring to the interaction between the \ac{VTwin} and \ac{PTwin}; (ii) support for concept drift detection to handle traffic variability; (iii) the presence of twin synchronization mechanisms, regardless of whether they explicitly use concept drift detection; and (iv) consideration of \ac{NDT} \ac{LCM} elements, such as the planning, training, or operational phases. The comparative analysis in Table~\ref{tabRelatedWork} emphasizes complementary and essential features of an \ac{NDT} considered in this work, rather than focusing solely on improving the \ac{VTwin} model itself, as is common in prior works. In this context, our objective is not to compare different \ac{VTwin} models but rather to consider a given \ac{VTwin} model and examine how our proposed architecture aims to preserve its predictive performance through a twin synchronization strategy. Finally, Table~\ref{tabRelatedWork} considers only studies that explicitly use the term \ac{NDT}. This decision reflects our intention to maintain terminological consistency. Although other platforms, such as \ac{CPS}~\cite{kathiravelu2019sd, Wu2021}, may share certain similarities with \acp{NDT}, they often involve broader specifications that extend beyond the scope of NDTs as assumed in our work.

\begin{table}[!ht]
	\caption{Comparison between \ac{NDT} characteristics of the related works and our proposed architecture with support for concept drift.}
	\centering
    \scalebox{1}{\begin{tabular}{lcccc}
  \toprule
  \textbf{Authors}                                                                                               & \textbf{Closed-loop}      & \textbf{\begin{tabular}[c]{@{}c@{}}Concept drift \\ detection support\end{tabular}} & \textbf{\begin{tabular}[c]{@{}c@{}} Twin \\ synchronization\end{tabular}} & \textbf{\begin{tabular}[c]{@{}c@{}}NDT LCM \\ elements\end{tabular}}\\ \midrule
  
  \begin{tabular}[c]{@{}c@{}}Rusek et al., 2020~\cite{rusek2020routenet}\end{tabular} & \xmark    & \xmark & \xmark & \begin{tabular}[c]{@{}c@{}}\xmark\end{tabular} \\ \midrule
  
  \begin{tabular}[c]{@{}c@{}}Happ, et al., 2021~\cite{happ2021graph}\end{tabular} & \xmark & \xmark & \xmark & \begin{tabular}[c]{@{}c@{}} \xmark\end{tabular} \\ \midrule
  
  \begin{tabular}[c]{@{}c@{}}Wang et al., 2022~\cite{Wang2022}\end{tabular} & \xmark     & \xmark& \xmark & \begin{tabular}[c]{@{}c@{}}\xmark\end{tabular} \\ \midrule
  
  \begin{tabular}[c]{@{}c@{}}Güemes-Palau et al., 2022~\cite{guemes2022accelerating}\end{tabular} & \xmark & \xmark& \begin{tabular}[c]{@{}c@{}}\xmark\end{tabular}  & \begin{tabular}[c]{@{}c@{}}\xmark\end{tabular} \\ \midrule
  
  \begin{tabular}[c]{@{}c@{}}Ferriol-Galmés et al., 2022~\cite{ferriol2022flowdt}\end{tabular}    & \xmark     & \xmark & \xmark & \begin{tabular}[c]{@{}c@{}}\xmark\end{tabular}  \\ \midrule
  
  \begin{tabular}[c]{@{}c@{}}Saravanan et al., 2022~\cite{saravanan2022enabling}\end{tabular}     & \xmark     & \xmark & \xmark & \begin{tabular}[c]{@{}c@{}}\xmark\end{tabular} \\ \midrule
  
  \begin{tabular}[c]{@{}c@{}}Lai et al., 2023~\cite{lai2023deep}\end{tabular} & \xmark & \xmark & \begin{tabular}[c]{@{}c@{}}\xmark\end{tabular} & \begin{tabular}[c]{@{}c@{}}\xmark\end{tabular} \\ \midrule
  
  \begin{tabular}[c]{@{}c@{}}Farreras et al., 2023~\cite{Farreras2023}\end{tabular}               & \xmark     & \xmark & \xmark & \begin{tabular}[c]{@{}c@{}}\xmark\end{tabular} \\ \midrule
  
  \begin{tabular}[c]{@{}c@{}}Messaoudi et al., 2023~\cite{messaoudi2023}\end{tabular} & \Checkmark & \xmark & \xmark & \begin{tabular}[c]{@{}c@{}}\xmark\end{tabular}\\ \midrule
   
  \begin{tabular}[c]{@{}c@{}}Shin et al., 2023~\cite{shin2023network}\end{tabular} & \xmark     & \xmark & \xmark & \begin{tabular}[c]{@{}c@{}}\Checkmark\end{tabular} \\ \midrule
  
  \begin{tabular}[c]{@{}c@{}}Yu et al., 2023~\cite{yu2023digital}\end{tabular} & \xmark     & \xmark & \xmark & \begin{tabular}[c]{@{}c@{}}\Checkmark\end{tabular} \\ \midrule
  
  \begin{tabular}[c]{@{}c@{}}Ferriol-Galmés et al., 2023~\cite{ferriol2023routenet}\end{tabular}  & \xmark     & \xmark & \xmark & \begin{tabular}[c]{@{}c@{}}\xmark\end{tabular} \\ \midrule
    \begin{tabular}[c]{@{}c@{}}Zalat et al., 2024~\cite{Zalat2024}\end{tabular} & \Checkmark     & \xmark & \Checkmark & \begin{tabular}[c]{@{}c@{}}\xmark\end{tabular} \\ 
  \midrule
  \begin{tabular}[c]{@{}c@{}}Modesto et al., 2024~\cite{Modesto2024}\end{tabular} & \xmark     & \xmark & \xmark & \begin{tabular}[c]{@{}c@{}}\xmark\end{tabular} \\ 
  \midrule

    \begin{tabular}[c]{@{}c@{}}Farreras et al., 2025~\cite{farreras2025gnnetslice}\end{tabular}     & \xmark     & \xmark & \xmark & \begin{tabular}[c]{@{}c@{}}\xmark\end{tabular} \\ \midrule   
    
  \begin{tabular}[c]{@{}c@{}}Aben-Athar et al., 2025~\cite{abenathar2025}\end{tabular}     & \Checkmark     & \xmark & \xmark & \begin{tabular}[c]{@{}c@{}}\xmark\end{tabular}  \\ \midrule
  
  \textbf{Our work} & \Checkmark & \Checkmark & \begin{tabular}[c]{@{}c@{}}\Checkmark\end{tabular} & \begin{tabular}[c]{@{}c@{}}\Checkmark\end{tabular} \\
  \bottomrule
  \end{tabular}}
  \label{tabRelatedWork}
\end{table}

\section{Enhanced Network Digital Twin}
\label{sec: proposed NDT}

\begin{figure*}[!h]
    \centering
    \includegraphics[scale=0.6]{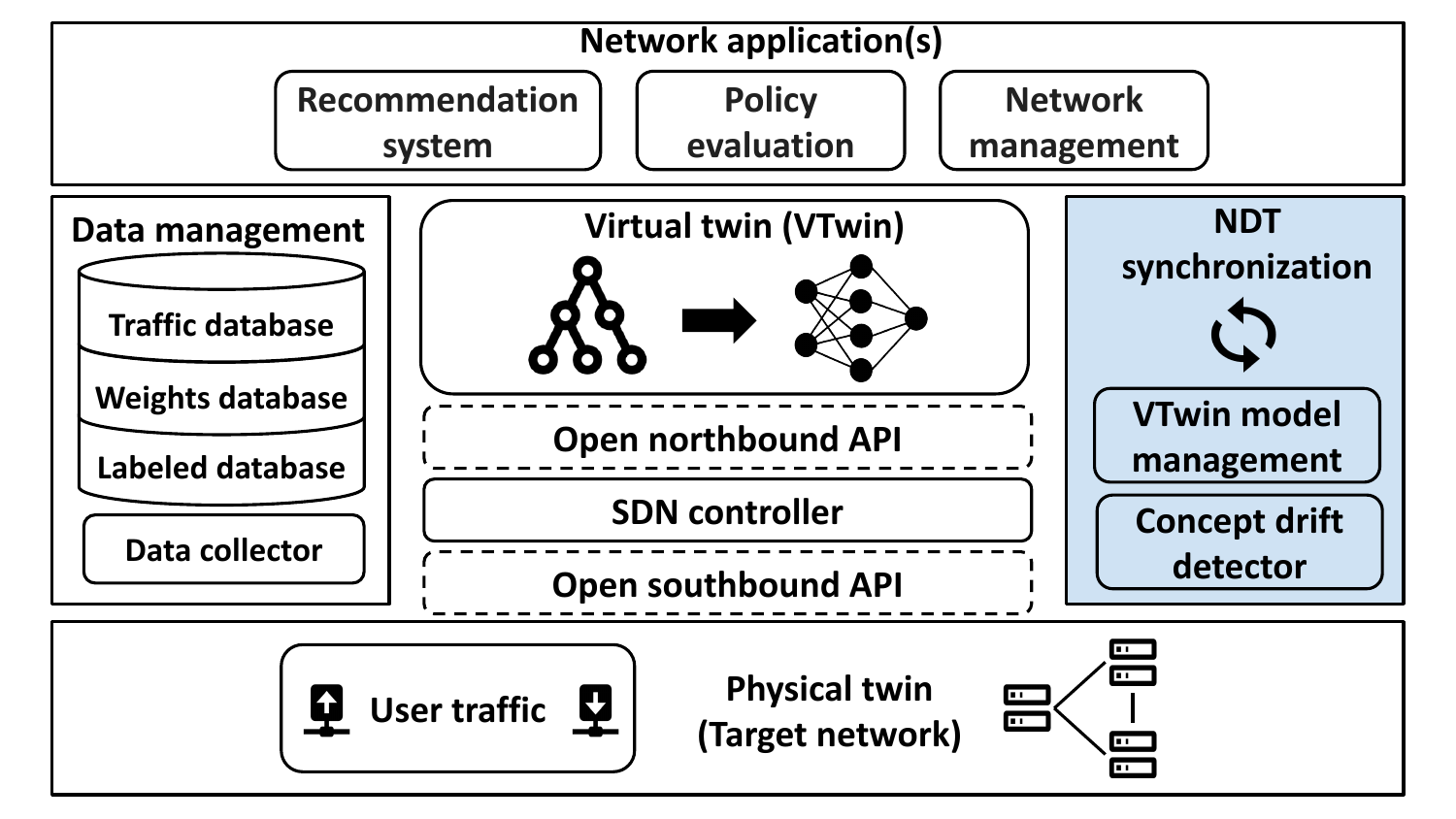}
    \caption{Enhanced \ac{NDT} architecture to construct a reactive approach against traffic variability. This self-adaptivity is achieved through the proposed implementation of the \ac{NDT} synchronization functions (in blue), including the concept drift detector and \ac{VTwin} model management.}
\label{fig:proposed_architecture}
\end{figure*}

The proposed enhanced architecture is divided into the modules illustrated in Figure~\ref{fig:proposed_architecture} and shares some inheritance from standardization bodies \cite{itu_digital_twin, ndt_ietf} and industry players~\cite{ohlen2022}. Following a bottom-up description, the first module corresponds to the \ac{PTwin} defined by the switches in the transport network. Above the \ac{PTwin}, we have the \ac{SDN} controller and its related interfaces, northbound and southbound. Next to it, we have the \ac{VTwin}, which in our solution is implemented using a \ac{GNN} model. It provides a digital representation of the relationship between network traffic and topology links. This model plays a crucial role in estimating the \ac{QoS} parameters in the \ac{PTwin}. Table~\ref{tab:paper_terms} provides an overview of the terms used to represent both twins. The following module is the data management system, which stores useful features of the traffic database (see Table \ref{tab:flow_traffic_features}). This module maintains a unified repository of transport network data, which is essential for feeding the \ac{VTwin} with relevant features in the retraining process. Additionally, this module includes the necessary functions to collect and store traffic data, ensuring seamless integration with the overall system. 

\begin{table}[!h]
    \centering
    \caption{Description of terms used.}
    \begin{tabular}{l|l}
    \toprule
        Term & Description \\
    \midrule
         $\mathcal{G}$ & Transport network topology \\
         $\mathcal{F}$ & Set of network flows \\
         $\mathcal{V}$ & Set of nodes \\
         $\mathcal{E}$ & Set of links \\
         $O$ & Source switch \\
         $D$ & Destination switch \\
         $e_j$ & $j$-th link \\
         $f_t$ & $t$-th flow \\
         $\mathbf{x}_{e_j}$ & $j$-th link feature vector \\
         $\mathbf{x}_{f_t}$ & $t$-th flow feature vector \\
         $x$ & Average traffic rate \\
         $T$ & Total number of flows \\
         $J$ & Total number of links in a path \\
         $P_t$ & Probability distribution at discrete time $t$ \\
         $a(\cdot)$ & Attention score \\
         $\psi(\cdot)$ & Message function \\
         $\phi(\cdot)$ & Update function \\
         $\theta$ & \Ac{MLP} learnable parameters \\
         $\bigoplus$ & Aggregation function \\
         $\mathcal{O}$ & Readout output \\
         $\mathbf{j}$ & Vector of link indices \\
         $\mathbf{i}$ & Vector of flow indices \\
         $K$ & Number of iteration on message passing \\
         $\mathcal{E}'$ & Path of a specific flow \\
         $p_j$ & j-th link propagation delay \\
         $q$ & Type of \ac{QoS} metric (delay or jitter) \\
         $y^{(q)}$ & per-flow \ac{QoS} metric \\
         $\hat{y}^{(q)}$ & Predicted per-flow \ac{QoS} metric \\
         $m$ & Flow embedding size \\
         $n$ & Link embedding size \\
         $\mathbf{h}$ & Embedding vector \\
         $\mathcal{N}$ & Neighborhood of a node \\
    \bottomrule
    \end{tabular}
    \label{tab:paper_terms}
\end{table}

Operating in the background to enable communication between these \ac{NDT} modules, we have the interfaces defined by the \ac{SDN} controller, which allow for indirect data exchange between the \ac{VTwin} and the controller. This setup is intended to apply any necessary modifications in the \ac{PTwin} and facilitate indirect data exchange between the database collector and the \ac{PTwin} to fulfill this database with updated information from the physical environment. With this closed loop of control and traffic data between both twins, the \ac{VTwin} predictions can be used in high-level applications, either with the involvement of a network operator, as defined by Aben-Athar et al.  \cite{abenathar2025}, or autonomously, without human supervision, as demonstrated by Messaoudi et al. \cite{messaoudi2023}. In this article, we assume an \ac{SLA} monitoring application. Complementary to this closed loop, the main contribution relies on the modules related to the synchronization of the \ac{NDT} modules, which are responsible for synchronizing the status of the transport network in both domains in a reactive manner. Along with the concept drift detector, this module monitors whether the incoming traffic that reaches the \ac{VTwin} remains similar. Once changed and detected by the drift detector, this module has the capability to trigger a retraining process, as this change can also mean that the \ac{VTwin} model is no longer predicting the total per-flow delay or jitter with acceptable error performance. Finally, the source code of the proposed closed-loop architecture, its integration with concept drift techniques, the datasets used, and the evaluation methodology are available in a public GitHub repository.\footnote{\url{https://github.com/lasseufpa/robust-ndt}}

\subsection{Physical twin}
As the \ac{PTwin}, we considered a transport network topology modeled as a 2-tuple graph $\mathcal{G} = (\mathcal{V}, \mathcal{E})$. In this graph, $\mathcal{V}$ denotes the set of nodes, each representing a network switch, while $\mathcal{E} \subseteq \mathcal{V} \times \mathcal{V}$ defines the set of edges (i.e., links) that connect pairs of nodes. Within this network, a traffic generator can produce a set of flow traffic instances denoted as $\mathcal{F} = \{f_1, f_2, \dots, f_t, \dots, f_T\}$, with a total of $T$ flows. Each flow instance consumes network resources, such as link bandwidth, by applying variable load intensities along a specific path composed of links $\mathcal{E}' = \{e_1, e_2, \dots, e_j, \dots, e_J\}$, where $J$ denotes the flow length. We also define each flow instance as a tuple $f_t = (\mathbf{x}_{f_t}, O, D, \mathbf{j})$, where $\mathbf{x}_{f_t} \in \mathbb{R}^4$ is a feature vector that characterizes the flow's behavior. In this sense, each feature vector $\mathbf{x}_{f_t}$ is associated with a mapping rule $h$ to a real number defined as the interested \ac{QoS} metric, such that $h: \mathbf{x}_{f_t}\rightarrow \{y^{(q)}\}$, where $q \in \{\text{delay}, \textrm{jitter}\}$, and  such that $y^\textrm{delay}$ and $y^\textrm{jitter}$ represent the per-flow delay and jitter, respectively. Furthermore, $O$ and $D$ represent the source and destination switches ($\{O, D\} \in \mathbb{Z}^+$); and $\mathbf{j}$ is a vector of link indices that the $t$-th flow traverses. Similarly, each link instance is defined as a tuple $e_j = (\mathbf{x}_{e_j}, \mathbf{i})$, where $\mathbf{x}_{e_j} \in \mathbb{R}^2$ is a feature vector describing the link’s characteristics, and $\mathbf{i}$ is a vector of flow indices representing the flows traversing that $j$-th link. The complete list of features for both flow and link instances is provided in Table \ref{tab:flow_traffic_features}. Among these features, special attention should be given to the propagation delay, which was converted from a link-level feature to a flow-level feature by summing the propagation delays \( p_j \) of the \( j \)-th links along the path $\mathcal{E}'$. Furthermore, the average traffic rate \( x \in \mathbf{x}_{f_t} \) at time \( t \) is modeled as a random variable following a probability distribution \( P_t \), such that \( x \sim P_t \).

\begin{table}[!h]
    \centering
    \caption{Flow and link features that are extracted to be used as input feature for training and testing the \ac{VTwin} model.}
    \scalebox{1}{\begin{tabular}{lcc}
    \toprule
       Feature  & Unit & Type \\
       \midrule
       Average traffic rate & \si{bits/s} & Flow \\
       Path propagation delay & \si{s} & Flow \\
       Flow length & --- & Flow \\
       Average packet loss & \si{packets/s} & Flow \\
       Capacity & \si{bits/s} & Link \\
       Link load & \% & Link \\
       \bottomrule
    \end{tabular}}
    \label{tab:flow_traffic_features}
\end{table}

\subsection{Database management and SDN controller}

\begin{figure}[!h]
    \centering
    \includegraphics[scale=0.48]{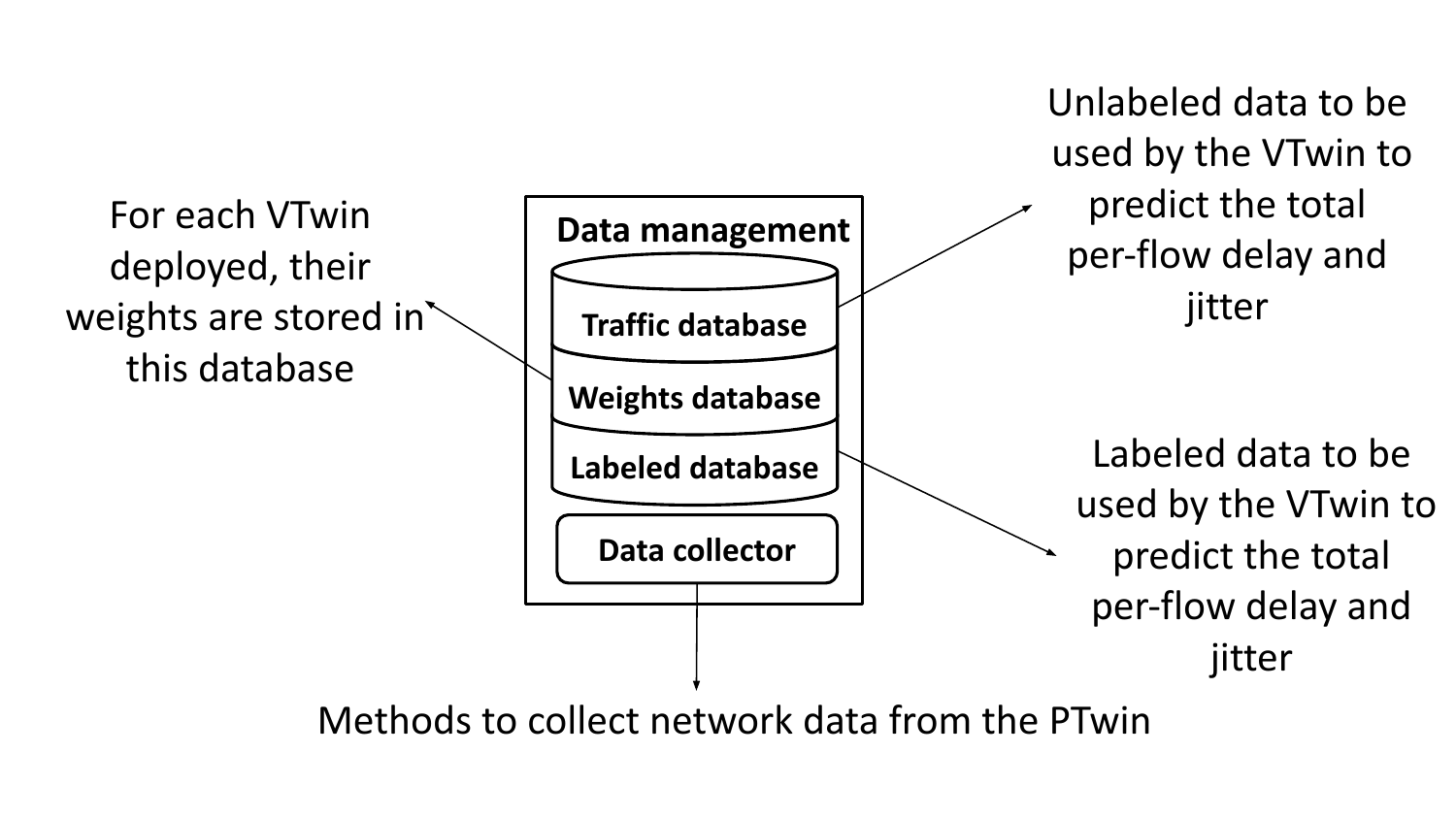}
        \caption{Primary roles of each database in the NDT data management module.}
    \label{fig:roles_database}
\end{figure}

The database used for the decision-making process is divided into three types of storage. The first one, called the \textit{traffic database}, is related to the data stream that was filtered after the traffic generator created it and will serve as input for the \ac{VTwin} model to perform the necessary \ac{QoS} prediction; i.e., the total per-flow delay or jitter labels are unavailable. The second type of database is related to the examples that should be used to retrain our \ac{VTwin} model, and it is called a \textit{labeled database}, as we manipulated the ground truth delay or jitter to update the \ac{VTwin} model weights when necessary. The last storage refers to the \textit{weights database}, which is aimed at storing the historical \ac{VTwin} model weights during the \ac{NDT} operation. In summary, the primary roles of each database are depicted in Figure~\ref{fig:roles_database}.

To store this raw data, we are also considering data collection functions to discard unnecessary packet-level traffic data, such as the \ac{IP} addresses of the server and client and transmission timestamps. Therefore, we aim to store only \ac{QoS} related features that we can extract from the \ac{PTwin} environment and that are useful for the \ac{VTwin} prediction task. It is important to note that we assume the data required for the retraining process has already been collected, labeled, and stored in our database. Furthermore, issues related to synthetic data generation or real-time data collection for retraining are considered outside the scope of this work. 

The interfaces employed in the proposed architecture (Figure  \ref{fig:proposed_architecture}) enable the indirect exchange of control data between both twins, with the \ac{SDN} controller serving as the network operating system~\cite{Kreutz2015}. This type of interaction is similar to that of a controller-based \ac{CPS}~\cite{molina2018software, salazar2019, kathiravelu2019sd}, which also leverages SDN controller interfaces to translate high-level network policies into control instructions executed at the physical layer. With this context, we utilize the \ac{SDN} controller as the central bridging tool, leveraging its available \ac{API}. Specifically, another feature stored in the traffic and labeled database consists of the vectors $\mathbf{i}$ and $\mathbf{j}$. Consequently, one of the critical functions of the interface module is to work with the data collection's path collector, which identifies and records the routes taken between each source–destination node pair. The path collector operates based on a specific node identifier, which is a 16-character hexadecimal string derived from the \ac{PTwin}. Therefore, given that flow traffic was initialized between two pairs of nodes, it is possible to consult the path of each flow through the controller \ac{REST} \ac{API}. In this case, it is assumed that the \ac{SDN} controller operates in a reactive forwarding manner; that is, it considers forwarding decisions and route establishment on demand, working only when a packet needs to be sent. Proactive forwarding can also be considered in this context by the \ac{SDN} controller, since this configuration is transparent to \ac{VTwin} (\ac{VTwin} only needs to know the path that a given flow traverses). Moreover, for different source destination switch pairs, we assume that the shortest path defines the routing table between them.  

\subsection{Virtual twin}

\begin{figure}[!h]
    \centering
    \includegraphics[scale=0.46]{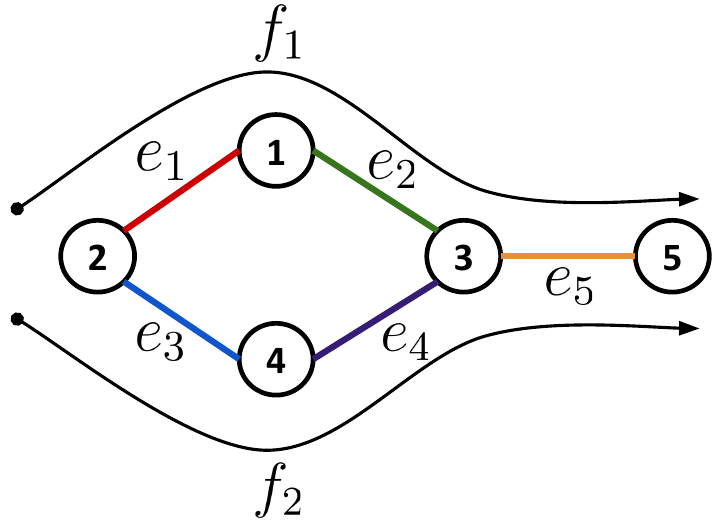}
    \caption{Links and flows relationship in a generic network topology.}
    \label{fig:f_to_l_l_to_f}
\end{figure}

\begin{figure}[!h]
    \centering
    \includegraphics[scale=0.48]{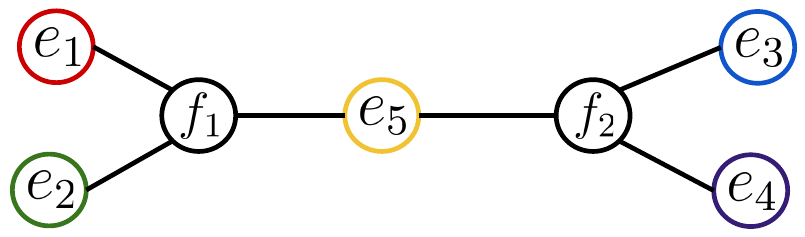}
    \caption{Heterogeneous graph, considering flow and link node types, generated from the transport network of Figure \ref{fig:f_to_l_l_to_f}.}
    \label{fig:heterogeneous_graph}
\end{figure}

Considering the adopted \ac{NDT} architecture, the \ac{VTwin} assumes a role in mimicking key aspects of its \ac{PTwin}. In this sense, the \ac{VTwin} is based on a \ac{GNN}, a type of deep learning model. Accordingly, the \ac{VTwin} is responsible for abstracting the most complex organization of the transport network and synthesizing its key aspects to evaluate the interest \ac{QoS} parameter while maintaining a certain level of fidelity in the physical-virtual relationship. For the objective of estimating the total per-flow delay and jitter, it is essential to model the relationship between the flow traffic applied to propagate information through the wired medium and the links used for this purpose, as shown in the generic graph in Figure \ref{fig:f_to_l_l_to_f}. This modeling step is fundamental, as the \ac{VTwin} input is not the original graph representing the transport network, but rather the hypergraph derived from it. The relationship between the flows and links of Figure \ref{fig:f_to_l_l_to_f} creates an essential circular dependency that is explored to generate the heterogeneous graph used as input for the \ac{VTwin} model. 
In this case, we can see that flow $f_1$ depends on links $e_1$, $e_2$, and $e_5$. Similarly, the flow $f_2$ has a dependency on the links $e_3$, $e_4$, and $e_5$. Therefore, these relationships generate the \ac{VTwin} input graph depicted in Figure \ref{fig:heterogeneous_graph}.

\begin{figure}[!h]
    \centering
    \includegraphics[scale=0.46]{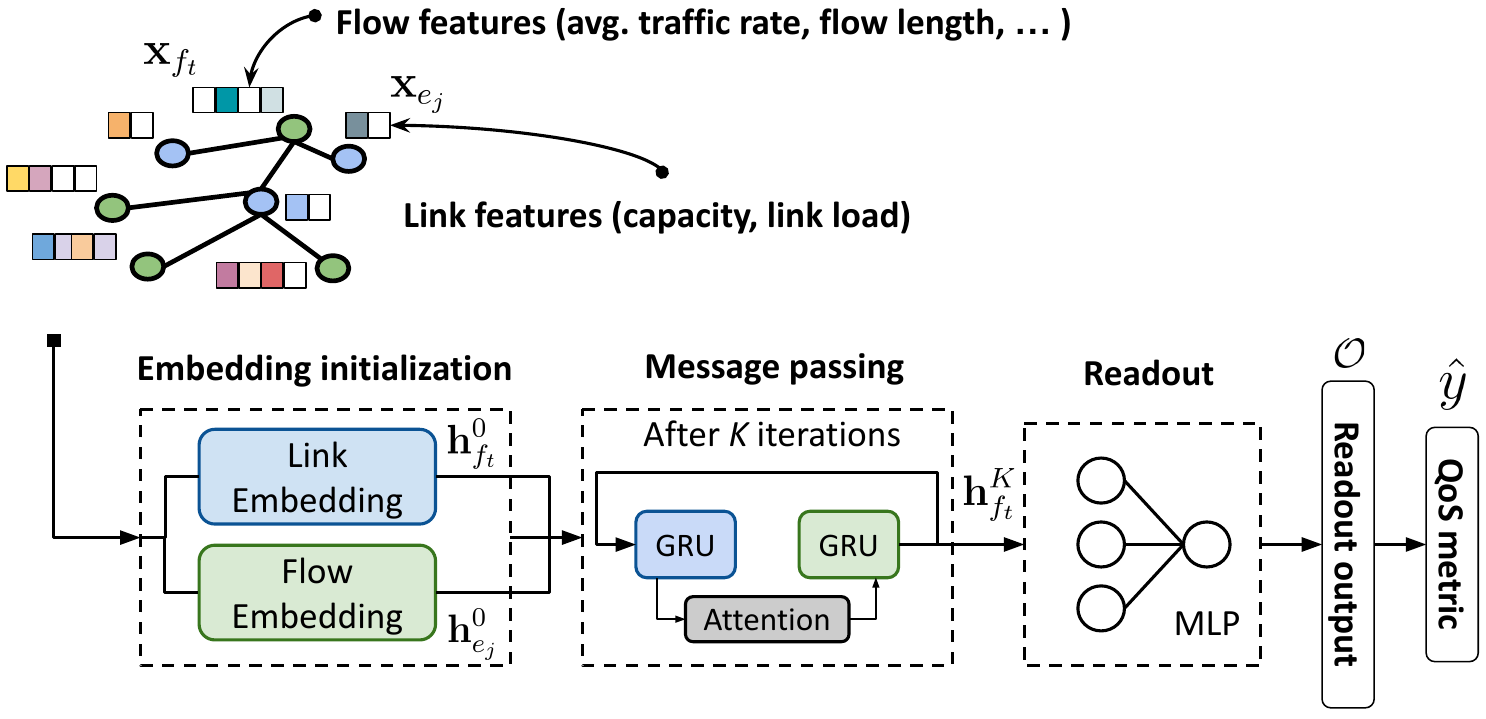}
    \caption{The adopted GNN architecture takes as input both flow and link features from the heterogeneous graph. It is composed of three modules: embedding initialization, message passing, and readout. Its output consists of QoS metrics such as delay and jitter.}
    \label{fig:gnn_workflow}
\end{figure}

Regarding the architecture of the \ac{VTwin}, we have a \ac{GNN} based on a modified RouteNet-Fermi with a self-attention mechanism \cite{Modesto2024}, as illustrated in Figure~\ref{fig:gnn_workflow}. This model processes flow and link characteristics through message passing across two \ac{GRU}-based layers. The aggregated messages are used by a readout function defined by a \ac{MLP} to compute the parameters used to estimate \ac{QoS} metrics.

The embedding initialization aims to process the node features $\mathbf{x}_{f_{t}}$ and $\mathbf{x}_{e_j}$ from a heterogeneous graph obtained from the relationship between the links and the flows; that is, the $f_{t}$ flows that passed through several links and the $e_j$ links that received each flow~\cite{Modesto2024}. Since this graph has two types of nodes with different feature vectors, specific embedding initialization functions are used to convert the features of each node type from their original dimensions to high-dimensional space $\mathbf{h}_{f_{t}} \in \mathbb{R}^m, \mathbf{h}_{e_j} \in \mathbb{R}^n$, where $m$ and $n$ are model hyperparameters, respectively, representing the sizes of the flow and link embedding vectors.

To obtain these high-dimensional vectors for each node type, two \ac{MLP} layers are used as an initialization function, both with one hidden layer and having the output dimension defining the embedding size, $\|\mathbf{h}_{f_{t}}\| = m$ and $\|\mathbf{h}_{e_j}\| = n$. Thereby, this embedding initialization is defined by equations:
\begin{equation}
    \mathbf{h}_{f_t} = \text{MLP}\left(\mathbf{x}_{f_{t}}, \theta_{f_{t}}\right),
\end{equation}
and
\begin{equation}
    \mathbf{h}_{e_j} = \text{MLP}\left(\mathbf{x}_{e_j}, \theta_{e_j}\right),
\end{equation}
\noindent where $\theta_{f_t}$, $\theta_{e_j}$ are the learnable parameters of each initialization function. 

In the message-passing phase, these high-dimensional vectors are processed iteratively during $k \le K$ (which $K$ is another model hyperparameter) time steps by the message function $\psi(\cdot)$ and updated by functions $\phi_\textrm{f}(\cdot)$, $\phi_\textrm{l} (\cdot)$, as defined as follows~\cite{Bronstein2021GeometricDL}:
\begin{equation}
    \mathbf{h}_{f_t}^{k+1} = \phi_\textrm{f}\left(\mathbf{h}_{f_{t}}^k,  \psi\left(\mathbf{h}_{f_{t}}^k, \mathbf{h}_{e_j}^k\right)\right),
\label{eq:mp_flow_to_link}
\end{equation}
and
\begin{equation}
    \mathbf{h}_{e_j}^{k+1} = \phi_\textrm{l}\left(\mathbf{h}_{e_j}^k, \bigoplus_{p \in \mathcal{N}_e}a\left(\mathbf{h}_{e_j}^k, \mathbf{h}_p^{k+1}\right)  \psi\left(\mathbf{h}_{e_j}^k, \mathbf{h}_p^{k+1}\right)\right),
\label{eq:mp_link_to_flow}
\end{equation}
\noindent where $\bigoplus$ is an aggregation function, $a(\cdot)$ is the attention score, and $\mathcal{N}_e$ is the neighborhood of the $e$-th link node. Moreover, this model treats the message function as a \ac{GRU} layer, and the auxiliary assignment operation as an update function. The output of the message-passing is used by the readout function to calculate the parameter $\mathcal{O}$, implemented using an \ac{MLP} with two hidden layers. This MLP produces a single-neuron output, as defined by the following equation:
\begin{equation}
    \mathcal{O} = \text{MLP}\left(\mathbf{h}_{f_t}^{k+1}, \theta_\textrm{o}\right),
    \label{eq: readout}
\end{equation}
\noindent where $\theta_\textrm{o}$ are the learnable parameters of the readout \ac{MLP} layer. 

Considering that a flow set passes through $J$ links, in the delay prediction task, this parameter $\mathcal{O}$ can be interpreted as queue occupancy, such that the predicted total per-flow delay $\hat{y}^{(\text{delay})}$ is evaluated by: 
\begin{equation}
\hat{y}^{(\textrm{delay})} = \underbrace{\sum_{j=0}^J 
\left(
  \dfrac{\mathcal{O}_j}{\mathcal{E}_j'}\right)}_{\text{Queue delay}} 
  \hspace{0.15cm}+ \underbrace{\sum_{j=0}^J p_j.}_{\text{Propagation delay}}
    \label{eq:delay_model}
\end{equation}

Although no explicit variations in queue configuration (such as buffer size) are introduced, the predicted total per-flow delay $\hat{y}^{\text{delay}}$ in Eq.~(\ref{eq:delay_model}) inherently includes the effect of queuing delay, derived using the output from Eq.~(\ref{eq: readout}), as the queue states implicitly affect each flow’s state~\cite{ferriol2023routenet}.

Finally, the predicted total per-flow jitter is directly evaluated using the output of readout function, as defined by:

\begin{equation}
    \hat{y}^{(\textrm{jitter})} = \sum_{j=0}^J\left(\dfrac{\mathcal{O}_j}{\mathcal{E}_j'}\right).
    \label{eq:jitter_model}
\end{equation}

\subsection{NDT Synchronization}
\label{sec:lcm_cd}

\begin{figure}[!h]
    \centering
    \includegraphics[scale=0.63]{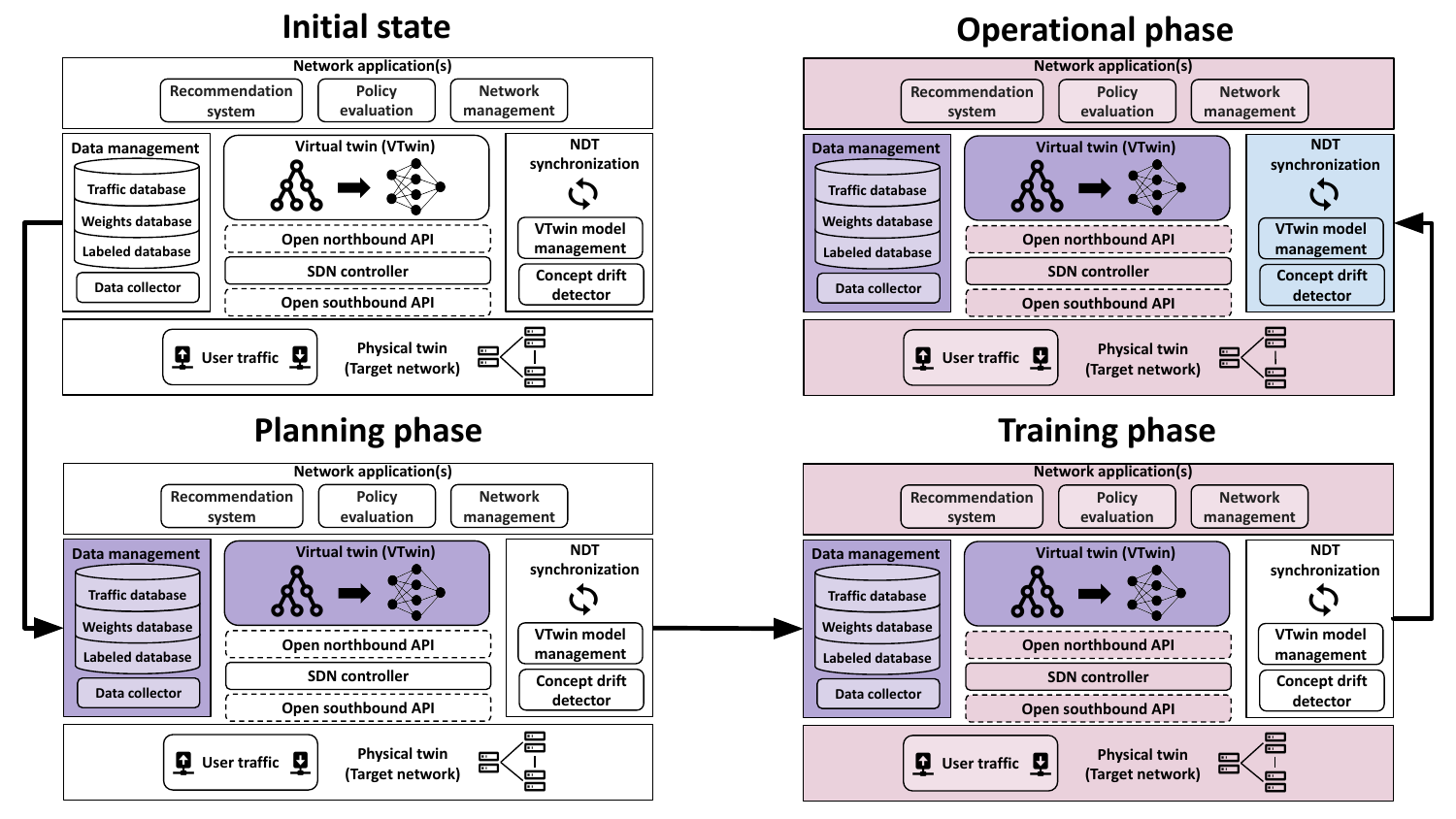}
    \caption{The \ac{NDT} \ac{LCM} comprises three phases: planning, training, and operational. Initially, no module is activated. During the planning phase, only the data management and \ac{VTwin} modules are enabled (purple blocks). In the training phase, interaction with the \ac{PTwin} by the \ac{SDN} interfaces and the use of a network application are required (light magenta blocks). Finally, in the operational phase, all modules are activated, and \ac{NDT} synchronization (light blue block) operates to maintain adequately the performance of the \ac{VTwin}.}
    \label{fig:lcm_phases}
\end{figure}

A well-defined \ac{LCM} provides the \ac{NDT} platform with the essential capabilities to function effectively in a production environment. It defines how the \ac{NDT} should behave under various conditions, especially during the operational phase \cite{Kherbache2021}, whether those conditions are routine or exceptional. Consequently, all \ac{LCM} phases can encompass different modules of the architecture presented in Figure \ref{fig:proposed_architecture}, each of which has a specific role and lifespan to ensure the correct operation of the \ac{NDT} tasks \cite{Kherbache2021}. Since each phase has a different lifespan, some modules shown in Figure~\ref{fig:proposed_architecture} may not exist in particular phases, as depicted in Figure~\ref{fig:lcm_phases}. 

For instance, during the planning stage of the \ac{VTwin}, the \ac{PTwin} is not yet required. At this point, the \ac{VTwin} can be trained using pre-existing datasets~\cite{bariah2024interplay}, considering related features from the target network environment and eliminating the need for real-time data integration. As a result, only the \ac{VTwin} and selected components of the data management module are active. This phase closely resembles the process of training an \ac{AI} model for predictive tasks using offline data, where the focus is on optimizing model parameters and hyperparameters based on generalization performance. Noteworthy is that the early stages of \ac{NDT} for the transport network, as seen in the related works, had a specific focus during this phase. As the \ac{NDT} lifecycle progresses, the training phase begins to incorporate the physical components of the \ac{SDN} infrastructure, including the interfaces connected to the \ac{SDN} controller. In this context, training refers to the process of evaluating the \ac{VTwin} prior to its deployment in a production environment. This phase enables the identification and mitigation of potential critical issues that could impact the proper functioning of the \ac{NDT} platform. Finally, in the operational phase, we employ two submodules that monitor the synchronization states of the \ac{NDT}: the concept drift detector and \ac{VTwin} model management. These submodules are essential as they ensure the \ac{NDT} robustness against traffic variability after \ac{NDT} planning and deployment, which is intimately associated with the degree of synchronization between the \ac{VTwin} and \ac{PTwin}~\cite{Alghamdi2024, liu2024digital}. This twinning rate will determine the frequency at which the \ac{NDT} will start the process of synchronizing the network states of the \ac{VTwin}~\cite{jones2020}. 

\subsubsection*{Concept drift detector}

\begin{figure}[!h]
    \begin{subfigure}[b]{0.48\textwidth}
    \centering
    \caption{Concept drift based on the changes of average traffic rate}
    \includegraphics[scale=0.42]{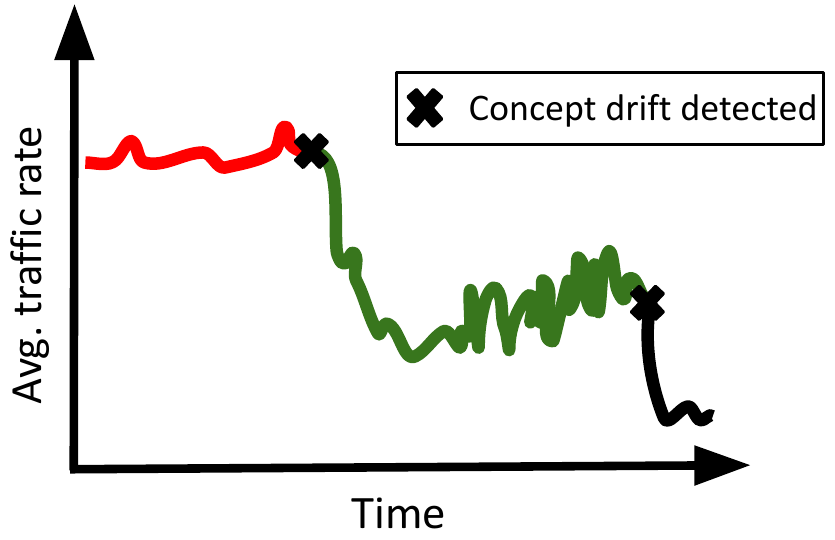}
    \label{fig:input_based_drift}
    \end{subfigure}    
    \begin{subfigure}[b]{0.5\textwidth}
    \centering
    \caption{Concept drift based on the changes of per-flow delay or jitter}
    \includegraphics[scale=0.414]{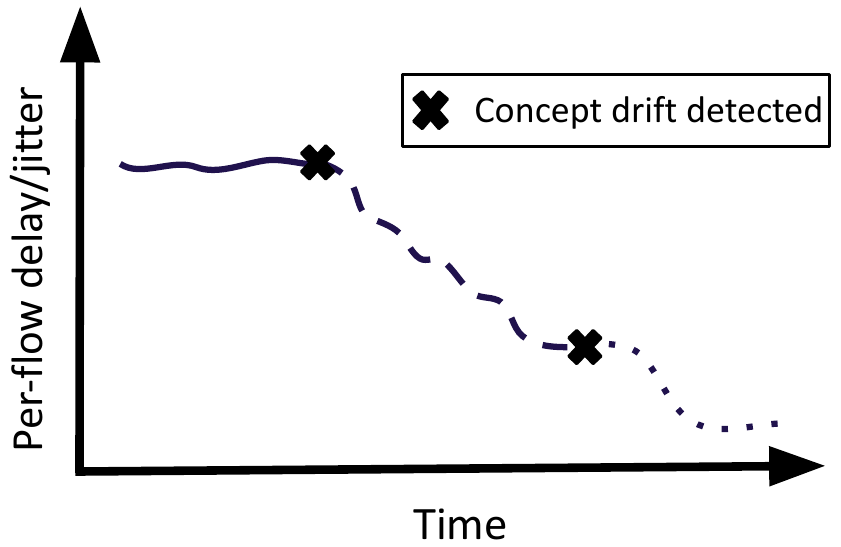}
    \label{fig:output_based_drift}
    \end{subfigure}
\caption{Types of data sources that can trigger a retraining process in a data-driven \ac{NDT}.}
    
\end{figure}

In a data-driven supervised learning approach, the synchronization process is associated with a retraining operation~\cite{Bagui2025} that can be triggered based on the input or output of the neural network model, as depicted in Figure~\ref{fig:input_based_drift} and \ref{fig:output_based_drift}, respectively. In this context, concept drift can be detected using input features such as the average traffic rate or output features such as per-flow delay and jitter. These features may be analyzed either individually or jointly by the concept drift detector. When considered jointly, the detection relies on a multivariate concept drift detector~\cite{lukats2025benchmark}, whereas analyzing each feature independently corresponds to a univariate approach.

\begin{figure}[!h]
    \centering
    \includegraphics[scale=0.43]{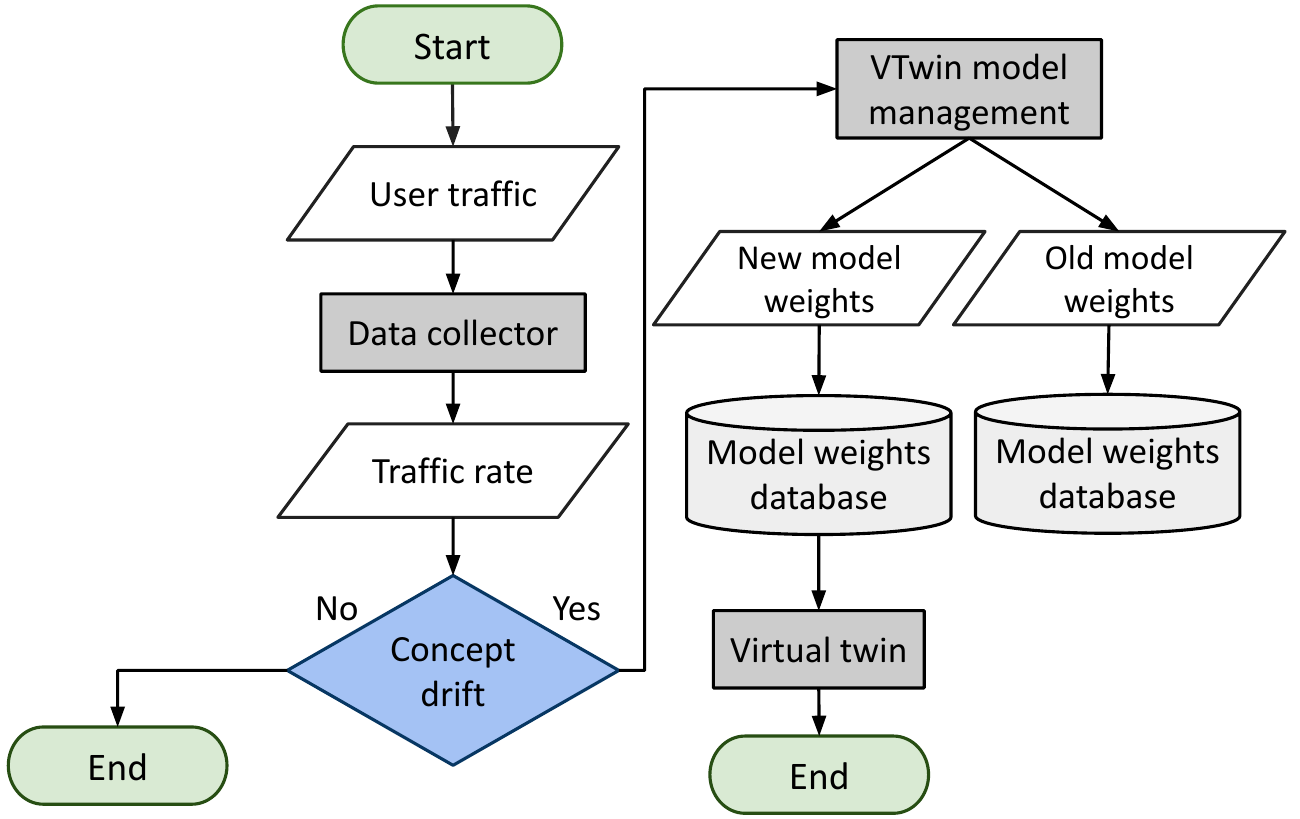}
    \caption{Sequence of steps in the operational phase to synchronize the \ac{VTwin} states based on the current \ac{PTwin}.}
    \label{fig:ndt_sync_fluxogram}
\end{figure}

This choice is essential, as the synchronization between the \ac{VTwin} and \ac{PTwin} will differ significantly depending on the target variable considered in the concept drift detection process. In the first case, synchronization is driven by the distribution pattern of each input feature, such as the network traffic of a flow passing through a set of links. In this sense, when decisions regarding concept drift are made solely based on the input data, the approach is referred to as a \textit{distribution-based} method~\cite{Bayram2022}. In contrast, when the correct labels are also considered, the method is classified as \textit{performance-based}~\cite{Bayram2022}. In the latter approach, within the context of this problem, synchronization depends on the label, specifically on the total per-flow delay or jitter. However, using this approach under realistic conditions, detecting the \textit{aging} of a neural network model is not straightforward, as the ground truth of the \ac{QoS} metric is not readily available for direct comparison with \ac{VTwin}'s predictions. Therefore, our synchronization approach employs a univariate distribution-based method that monitors the average flow traffic rate to detect concept drift during the \ac{VTwin} operation, thereby eliminating the need to rely on ground truth delay or jitter to trigger the synchronization process through \ac{VTwin} retraining. In operational terms, to reach these synchronization steps, we need to pass through different \ac{NDT} modules, as defined by the flowchart in Figure \ref{fig:ndt_sync_fluxogram}. 

The process starts with user traffic originating from the transport network. This raw traffic data is captured by the relevant network functions and stored in the traffic database. The data collector then extracts a key feature for use by the concept drift detector: the average traffic rate of each flow. This feature is selected because it effectively captures the primary characteristic of user traffic: its average throughput. Thereby, using a univariate covariate-based method as a concept drift detector, we can identify changes in the input distribution of features over time when \( P_t(x_{f_t}) \neq P_{t+w}(x_{f_{t+w}}) \), where \( t \) represents a given time instant, $\{x_{f_t} \in \mathbf{x}_{f_t}, x_{f_{t+w}} \in \mathbf{x}_{f_{t+w}} \}$ is a feature in time instant $t$ and $t+w$, in this case, the average traffic rate, and \( w \) are the sizes of the time window. In this context, our concept drift detector serves as a monitoring agent that detects changes in traffic patterns as incoming traffic is processed. This process allows us to indirectly estimate when the total \ac{QoS} metric distribution will begin to shift, signaling the expected degradation in \ac{VTwin} performance and triggering retraining.

\subsubsection*{VTwin model management}

When a concept drift is detected, the concept drift detector notifies the \ac{VTwin} model management submodule with a warning to start retraining using data that is already available in the labeled database. Upon the start of the retraining process, the \ac{VTwin} model management locks the trigger to prevent the drift detector from issuing duplicate warnings for the same drift event that has been previously addressed until the training is finished. Upon completion of the retraining process, the \ac{VTwin} model management sends a control message to the \ac{VTwin}, archives the previously trained model weights in a dedicated database, and updates the \ac{VTwin} model with the newly learned weights.

\subsection{Network application}

The network application is organized to use the predictions provided to the \ac{VTwin} to perform adjustments in the \ac{PTwin}, considering applications that may change the state of the \ac{PTwin}, such as routing optimization or flow admission control, as well as those related to the management of the network, such as \ac{SLA} monitoring for \textit{what-if} analysis~\cite{Wang2022} or emergency preparedness~\cite{TR28915}. To achieve this, various submodules can be employed within these applications, including, but not limited to, a recommendation system to propose different configurations in the \ac{PTwin}, a policy evaluation module to evaluate the configurations suggested by the recommendation system, and a network management submodule designed to provide application-level metrics from the \ac{PTwin}.

\subsubsection*{Recommendation system}
In scenarios where the network application aims to optimize certain states of the \ac{PTwin}, this module incorporates auxiliary tools, such as a recommendation system, that assist the network operator by leveraging outputs from the \ac{VTwin}. This recommendation system can generate new policies that can potentially be applied to the \ac{PTwin}. For example, in routing optimization, the recommendation system, which can be based on a genetic algorithm~\cite{Zalat2024} or a reinforcement learning model~\cite{abenathar2025}, may propose new routing policies based on total per-flow delay metrics provided by the \ac{VTwin}, aimed at optimizing the \ac{QoS} metrics on the transport network. Once suggested, this recommendation is sent to the policy evaluation submodule to be evaluated by a network operator if a human-in-the-loop \ac{NDT} is being used~\cite{ohlen2022, abenathar2025}.

\subsubsection*{Policy evaluation}
In a man-in-the-loop architecture, a policy evaluation submodule is essential to prevent network modifications without human oversight, which is an important safeguard in production environments, particularly in architectures built on \ac{5G} networks~\cite{messaoudi2023}. This is especially critical given the interdependence of network components, such as the reliance of the \ac{RAN} on the transport network. The role of this submodule is to consolidate all relevant information and assess the potential impacts on the network, enabling informed decision-making by the operator. Once a policy is evaluated, the final decision on its implementation rests with the operator. If approved, the corresponding control data is transmitted to the \ac{SDN} controller via the northbound API and subsequently applied to the PTwin, thus completing the \ac{NDT} loop.

\subsubsection*{Network management}
When the application serves as a network management tool for \textit{what-if} analysis, the task becomes conceptually simpler, as it does not require direct interaction with the \ac{PTwin} to implement changes. Instead, its main objective is to provide the network operator with insight into the possible outcomes of specific scenarios or conditions~\cite{TR28915}, which can be useful in network planning tasks. This work explores a particular use case in the Experimental Results section, focusing exclusively on the network management submodule to provide \ac{SLA} violations in the network~\cite{Sami2022, sommers2010}. The selected application assesses compliance with flow \acp{SLA}, specifically detecting violations related to the \ac{PDB} assigned to each packet. For example, if the total per-flow delay exceeds the \ac{PDB} assigned for this flow. So, for a given set of flows that traverse the transport network, this application can estimate the number of SLA violations, relying on the total per-flow delay predictions provided by the \ac{VTwin}.

\section{Experiments}
\label{sec:experiments}

\begin{figure}[!h]
    \begin{subfigure}[b]{0.48\textwidth}
    \centering
    \caption{Histogram of link capacities.}
    \includegraphics[scale=0.47]{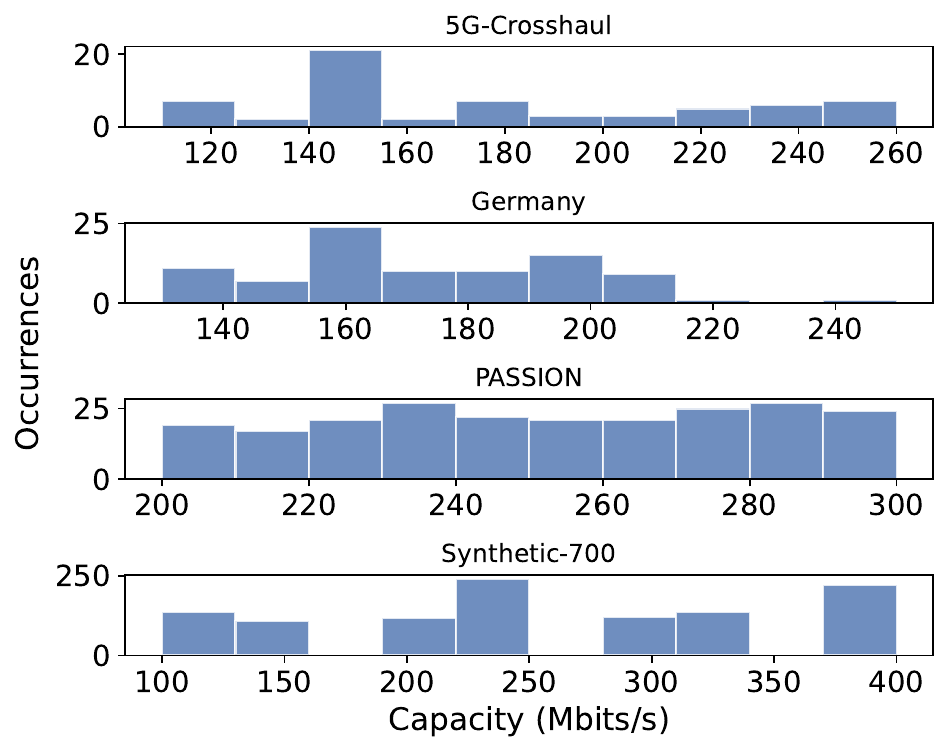}
    \label{fig:hist_capacity}
    \end{subfigure}    
    \begin{subfigure}[b]{0.5\textwidth}
    \centering
    \caption{Histogram of link propagation delay.}
    \includegraphics[scale=0.47]{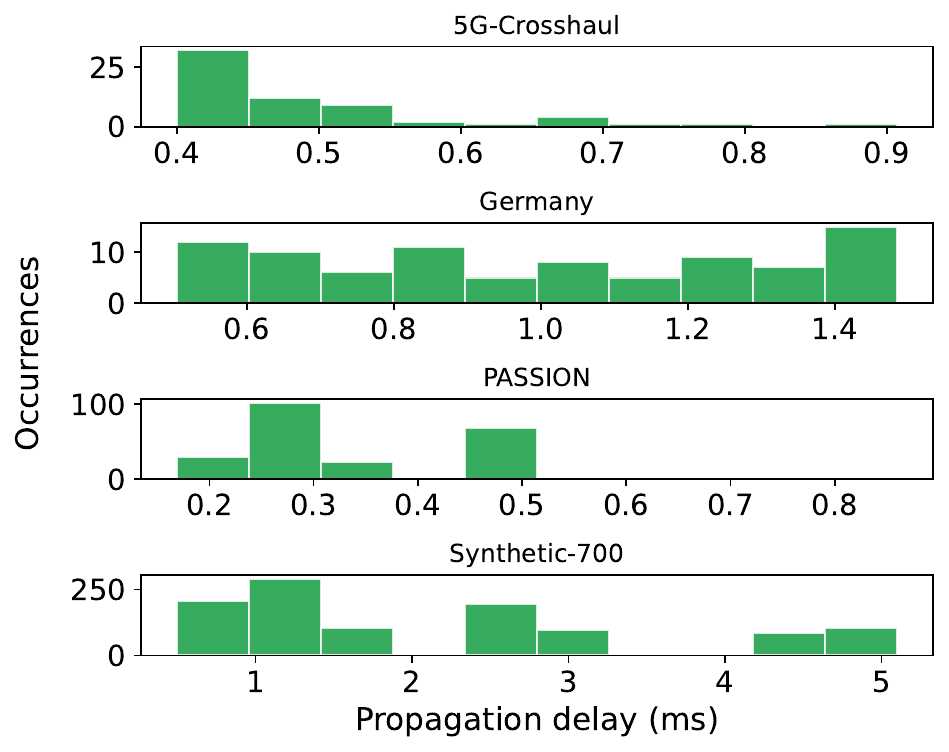}
    \label{fig:hist_propag_delay}
    \end{subfigure}

\caption{Link characteristics of the topologies used in our experiments.}
\end{figure}

The \ac{PTwin} integrates both emulation and simulation components used as part of the experiments to evaluate the proposed architecture. Using Mininet with the \ac{ONOS} controller, the emulated component focuses on topology organization, while the simulated component handles traffic generation. For emulated network topologies, we use four transport networks: the 5G-Crosshaul topology~\cite{Morais2023}, the PASSION project topology~\cite{Morais2023}, the \textit{Germany} topology~\cite{kdn_datasets}, and the Synthetic-700 topology comprising networks with 128, 51, 50, and 700 nodes, respectively, each with 63, 88, 224, and 1083 links. The details about the characteristics of link capacity and link propagation delay are also described in Figures~\ref{fig:hist_capacity} and ~\ref{fig:hist_propag_delay}, respectively. 

We employ the \ac{D-ITG} using a client-server model over the \ac{TCP} and \ac{UDP} protocols to generate realistic traffic patterns. To do so, we used a computer equipped with an Intel\textsuperscript{\textregistered}~Core\textsuperscript{\texttrademark}~i7-10700F CPU @ 2.90 GHz, an NVIDIA\textsuperscript{\textregistered} GeForce RTX 3060, 128 GB of RAM, and a storage capacity of 9 TB. This setup enables the incorporation of various traffic distributions within an emulation environment that is orchestrated using a suite of tools summarized in Table~\ref{tab:emulation_setup} to ensure accurate and reproducible testing conditions. 

\begin{table}[!h]
    \centering
    \caption{Setup configuration used to build our \ac{PTwin}.}
    \scalebox{0.95}{\begin{tabular}{lc}
    \toprule
    Tools & Version \\ 
    \midrule
      Operating system & Ubuntu 20.04 \\
      ONOS & 2.6.0 \\
      Mininet & 2.3.0\\
      Open vSwitch & 2.13.8 \\
      OpenFlow & 1.0 \\
      D-ITG & 2.8.1 \\
    \bottomrule
        \end{tabular}}
    \label{tab:emulation_setup}
\end{table}

We consider different distributions of packet size (in bytes) and traffic rate (in \si{Mbits/s}), without loss of generality, used to model internet traffic~\cite{Jurkiewicz2021, Song2005}, such as \textit{Poisson}, \textit{Exponential}, \textit{Uniform}, and Deterministic, in the presence of congestion. The parameters for each are defined in Table \ref{tab:networks_traffic_distro}, considering the minimum and maximum values $\mathcal{U}(512, 1024)$ for the uniform distribution, the mean for the exponential $\mathcal{E}(1024)$ and $\mathcal{P}(2048)$ for a Poisson distribution, as well as a deterministic pattern with traffic congestion defined by a constant value $K=512$ for a constant packet size in traffic generation. For the packet rate, we considered the same distribution as defined for the packet size. This simplification regarding the packet rate distribution was made to maintain a controlled environment and to gain a clearer understanding of the strengths and limitations of the proposed architecture.

\begin{table}[!h]
    \centering
    \caption{Packet size distributions used to simulate user traffic.}
    \scalebox{0.95}{\begin{tabular}{lcc}
    \toprule
       \makecell{Traffic pattern}  & Protocol &\makecell{Packet size  \\ (bytes)} \\
    \midrule
       Uniform  & TCP & $\mathcal{U}(512, 1024)$ \\
       Exponential & TCP & $\mathcal{E}(1024)$ \\       
       Poisson & TCP & $\mathcal{P}(2048)$\\
       Deterministic & UDP  & $K=512$\\
       \bottomrule
    \end{tabular}}
    \label{tab:networks_traffic_distro}
\end{table}

The datasets created from this traffic generation for a possible retraining process are presented in Table \ref{tab:type_of_datasets}, characterized by the four distribution patterns of packet size and traffic rate, the type of topology, and the number of flows (training examples). The training examples in this case refer to the number of samples that are to be used as input-output pairs in the model retraining.

\begin{table}[!h]
\centering
\caption{Type of labeled datasets stored used to retraining the \ac{VTwin}.}
\label{tab:type_of_datasets}

\scalebox{0.9}{\begin{tabular}{lccc}
\toprule
\# & \makecell{Traffic pattern} & Topology  & \makecell{Number of flows ($T$)} \\
\midrule
\multirow{5}{*}{1} & \multirow{5}{*}{Exponential} & 5G-Crosshaul & $121\,400$\\ \cmidrule{3-4}
 & & Germany  & $129\,600$\\ \cmidrule{3-4}
 &  & PASSION & $129\,600$ \\
 \cmidrule{3-4}
 &  & Synthetic-700 & $170\,000$ \\
 \midrule
\multirow{5}{*}{2} & \multirow{5}{*}{Poisson} & 5G-Crosshaul & $113\,100$ \\ \cmidrule{3-4}
 &  & Germany & $127\,800$ \\ \cmidrule{3-4}
 &  & PASSION & $113\,400$ \\ 
 \cmidrule{3-4}
 &  & Synthetic-700 & $194\,400$ \\
 \midrule
\multirow{5}{*}{3} & \multirow{5}{*}{Uniform} & 5G-Crosshaul & $112\,500$\\ \cmidrule{3-4}
 &  & Germany & $121\,400$ \\ \cmidrule{3-4}
 &  & PASSION & $112\,000$ \\
 \cmidrule{3-4}
 &  & Synthetic-700 & $162\,000$ \\  
 \midrule
 \multirow{5}{*}{4} & \multirow{5}{*}{\makecell{Deterministic}} & 5G-Crosshaul & $100\,800$\\ \cmidrule{3-4}
  &  & Germany & $61\,200$ \\
  \cmidrule{3-4}
 &  & PASSION & $123\,200$ \\
  \cmidrule{3-4}
 &  & Synthetic-700 & $186\,100$ \\  
 
 \bottomrule
\end{tabular}}
\end{table}

Considering these emulated environments, we propose several experiments using topologies that experience a sudden concept drift in traffic characteristics. We introduced an artificial drift at specific time intervals by altering the packet size distribution. As a result, the traffic follows the sequence of packet size and traffic rate patterns described below:

\begin{center}
   Start $\rightarrow$ \fbox{\mbox{$\mathcal{E}$}} $\rightarrow$ \fbox{\mbox{$\mathcal{P}$}} 
    $\rightarrow$ \fbox{\mbox{$\mathcal{U}$}} $\rightarrow$ \fbox{\mbox{$K$}} $\rightarrow$ End 
\end{center}

For each topology, we analyzed different time intervals of network traffic. Specifically, the total times in seconds for 5G-Crosshaul, Germany, PASSION, and Synthetic-700 are $25\,100$ \si{s}, $24\,600$ \si{s}, $26\,700$ \si{s}, and 39\,700 \si{s}, respectively. Moreover, the experiment was carried out with the understanding that we initially trained the \ac{VTwin} model using traffic flows characterized by an exponential packet size distribution. The first abrupt drift is introduced by switching the packet size distribution to a Poisson model. This changing distribution behavior persists until the final phase, where the traffic pattern transitions to a constant packet size distribution, accompanied by the presence of traffic congestion. Finally, we considered the \ac{NMSE} as a metric to evaluate the \ac{VTwin} performance with and without \ac{NDT} synchronization throughout these concept drifts, defined as: 
\begin{equation}
    \text{NMSE}^{(q)} = 10\log_{10}\left\{\dfrac{\sum\limits_{t=1}^T\left(y_t^{(q)} - \hat{y}_t^{(q)}\right)^2}{\sum\limits_{t=1}^T \left(y_t^{(q)}\right)^2}\right\},
    \label{eq:nmse_equation}
\end{equation}
\noindent where $\hat{y}_t^{(q)}$ and $y_t^{(q)}$ denote, respectively, the $t$-th predicted and ground-truth \ac{QoS} metric among $T$ test samples. The average values were computed using a sliding window of 100 consecutive samples over time. Although other window sizes could be used, a 100-sample window offers a suitable trade-off: it provides a sufficiently narrow confidence interval for the mean while maintaining high \ac{NMSE} sensitivity to short-term variations, an aspect particularly relevant during abrupt concept drift. We apply this evaluation across 10 independent realizations, aimed at obtaining an average perspective across the \ac{NDT} operation in each network topology.

Using this environment, our experiments consider a distribution-based detector aimed at detecting changes in the input. We configured the concept drift detector using the \ac{KSWIN} method~\cite{raab2020}, as implemented by the \textit{river} library~\cite{montiel2021river}. To work correctly, this method requires three hyperparameters: the \textit{significance} level, which is a sensitive parameter that controls the probability of false positives and is represented by a value of 0.001; a window size; and a buffer for the most recent samples. The sizes of these components vary across different topologies and are determined empirically.

\begin{figure*}[!h]
    \centering
    \includegraphics[scale=0.4]{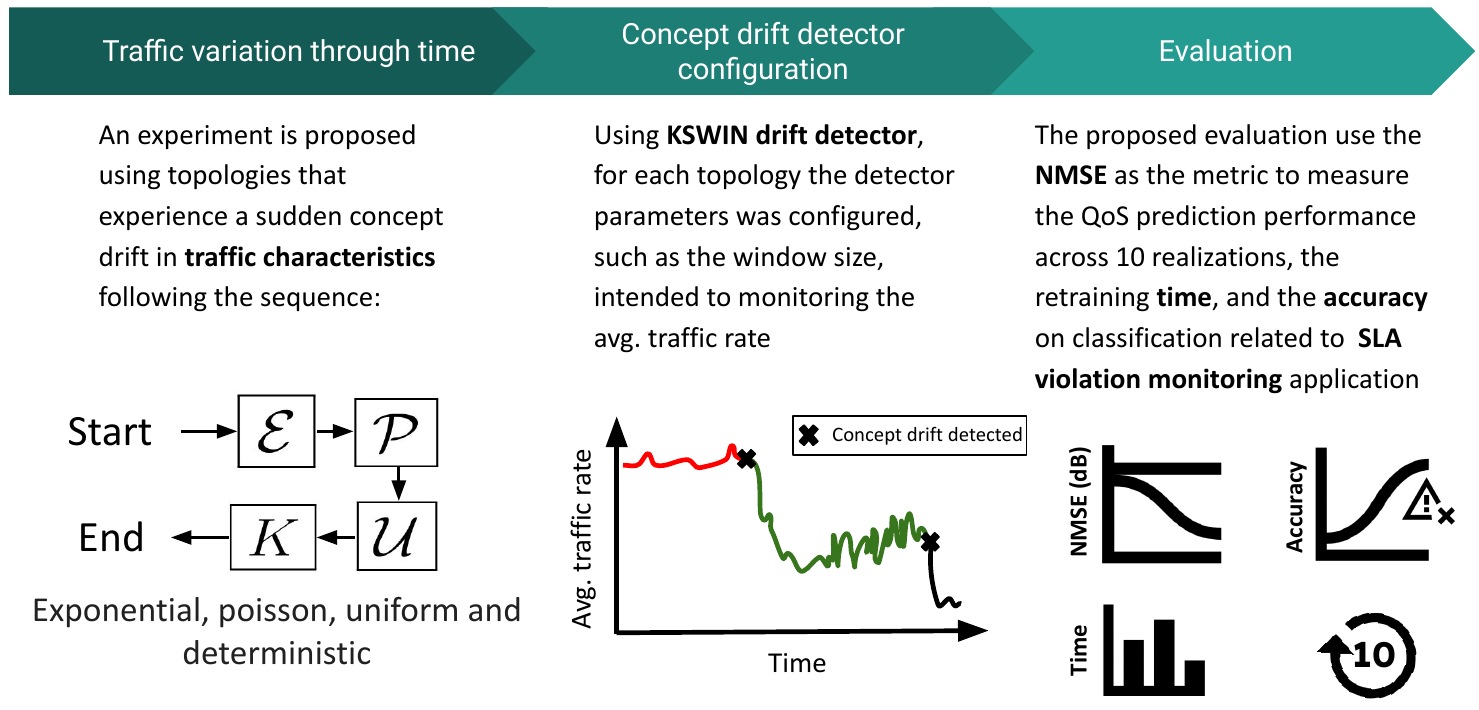}
    \caption{Experiment organization to evaluate the VTwin performance with and without synchronization in a scenario with concept drift inducted by different traffic pattern.}
    \label{fig:summary_evaluation}
\end{figure*}

For the \ac{VTwin} training, we used TensorFlow 2.14.0 and maintained consistency in all training parameters across all experiments. Specifically, we used a learning rate of $0.001$, 50 as the maximum number of epochs (\textit{early stopping} callbacks were also applied), applied Z-score normalization for feature scaling, adopted the \textit{Adam} optimizer, and used \ac{MAPE} as the loss function. Moreover, as the model hyperparameters, we used: $T = 8$ and $||\mathbf{h}_f|| = ||\mathbf{h}_l|| = 32$. This uniform configuration was chosen to ensure that any observed differences in model performance could be attributed solely to variations in the input data. In summary, our evaluation scenario can be described by Figure~\ref{fig:summary_evaluation}.

% \begin{table}[hpt]
%     \caption{\ac{VTwin} training parameters}
%     \centering
%     \scalebox{1.1}{\begin{tabular}{cc}
%        \toprule
%        Parameter  &  Used\\
%         \midrule
%         Learning rate & $0.001$ \\
%         Epochs & $50$ \\
%         Batch size & $100$ \\
%         Feature normalization & Z-score \\
%         Optimizer & \textit{Adam} \\
%         Loss function & \ac{MAPE} \\
%         \bottomrule
%     \end{tabular}}
%     \label{tab:training_parameters}
% \end{table}

\subsection*{Concept drift detector configuration}
\label{sec:results}

\begin{figure}[!h]
    \centering
    \includegraphics[scale=0.64]{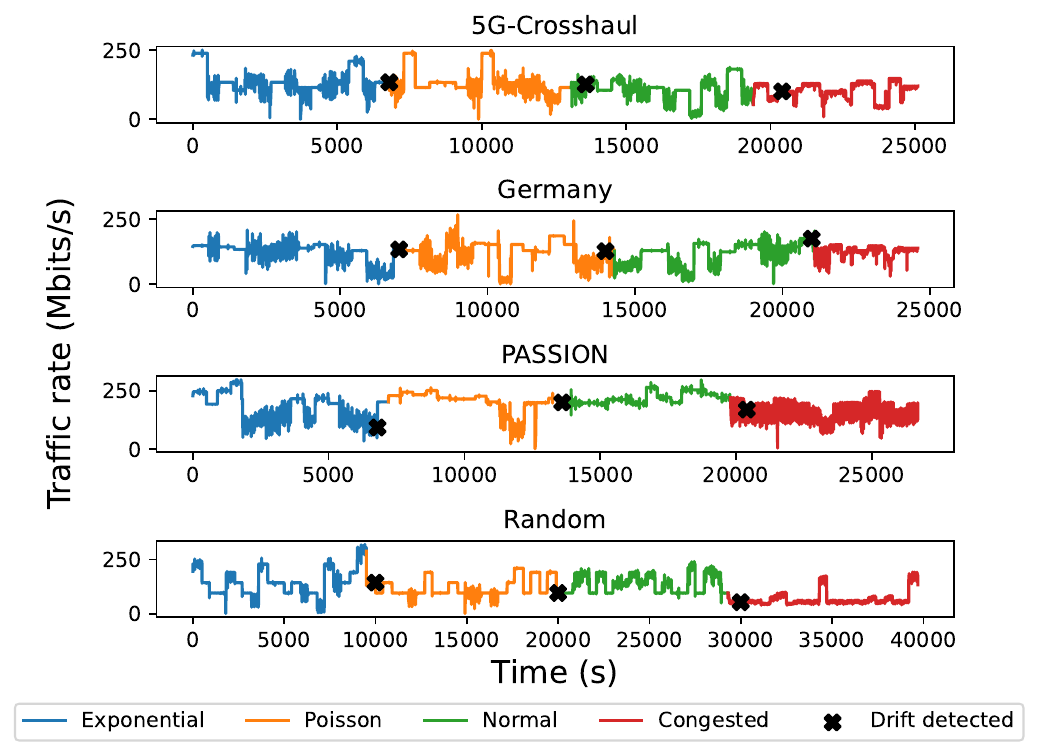}
    \caption{Drift detection pattern considering 5G-Crosshaul, Germany, PASSION, and Synthetic-700 topologies and four type of traffic patterns. Across the time series, each color in a given interval represents a different distribution pattern. The ``x'' points represent the detected concept drift.}
    \label{fig:drift_detection}
\end{figure}

We have the experiments from the \ac{KSWIN} results to evaluate the proposed synchronization elements, considering the four emulated topologies and the average traffic rate in \si{Mbits/s} as the features to be tracked. 
Figure~\ref{fig:drift_detection} illustrates the operation of the \ac{KSWIN} detector and the location of each detected drift. For each network topology, we selected a specific window size (in seconds): $6\,800$ for 5G-Crosshaul, $7\,000$ for Germany, $6\,800$ for PASSION, and $10\,000$ for the Synthetic-700 topology. Although alternative window sizes could be explored, determining the optimal size to minimize false positives falls outside the scope of this work.

\begin{figure}[!h]
    \centering
    \includegraphics[scale=0.45]{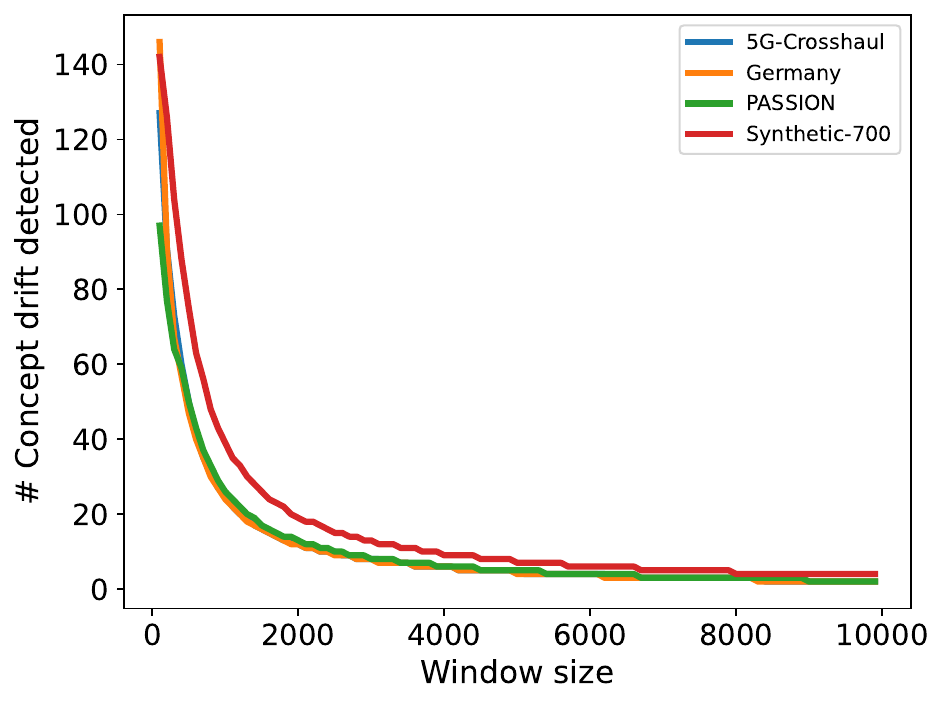}
    \caption{Relationship between the window size and the number of concept drift detection on the four topologies.}
    \label{fig:window_size_relation}
\end{figure}

In the evaluated environments, smaller window sizes generally result in a higher likelihood of false positives, as illustrated in Figure~\ref{fig:window_size_relation}, which analyzes various window sizes across all topologies. The curves in this figure exhibit a consistent pattern and converge in the number of detected concept drifts as the window size increases. %This behavior can be attributed to the timing of the concept drift events. For example, the first concept drift, shown in Figure~\ref{fig:drift_detection}, occurs approximately in second 7000 across 3 out of 4 topologies, coinciding with the point at which the curves in Figure~\ref{fig:window_size_relation} cease to converge. 

\subsection*{Delay prediction performance}

\begin{figure}[!h]
    \centering
    \includegraphics[scale=0.45]{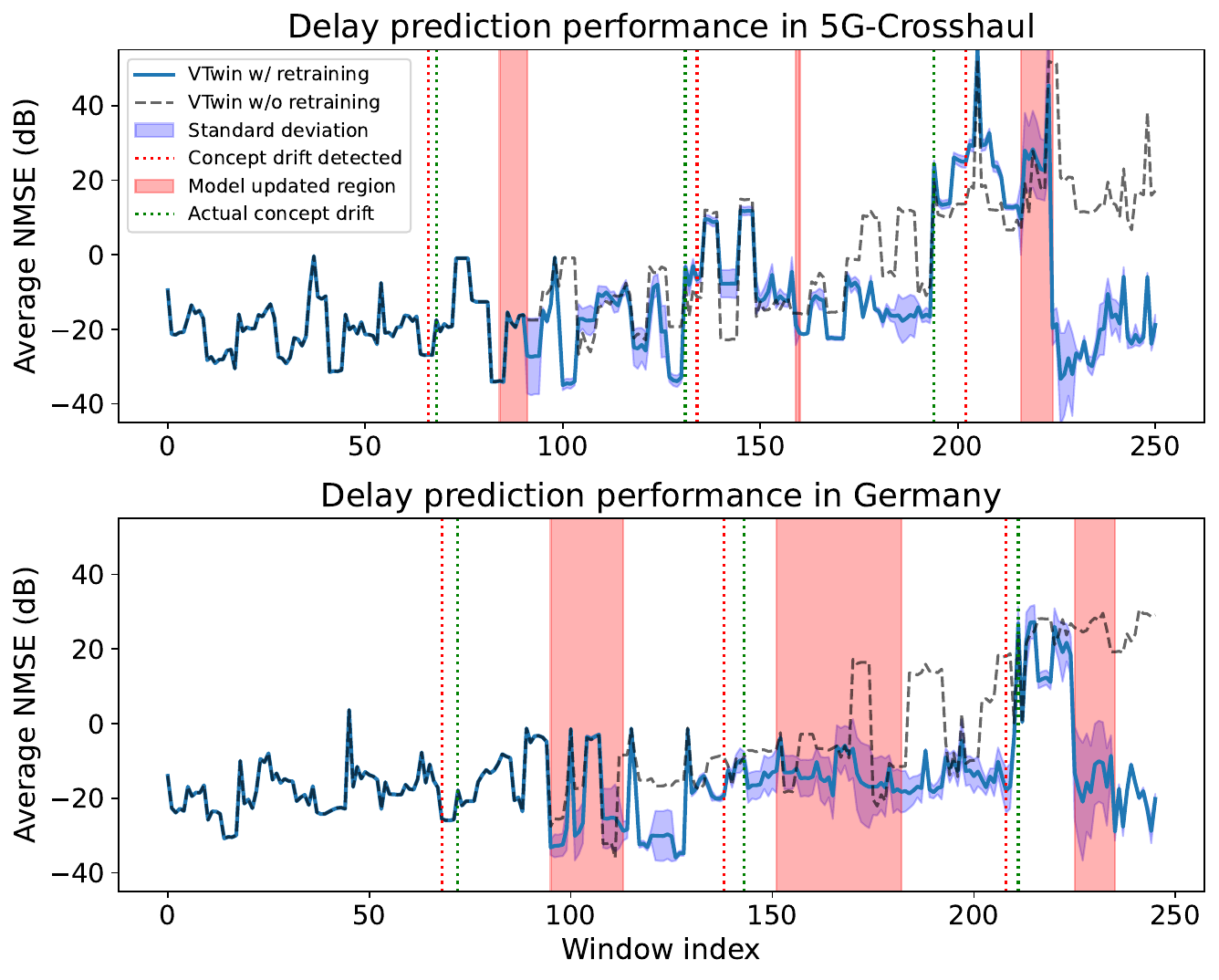}
    \caption{Validation of the proposed architecture in the 5G-Crosshaul and Germany topologies. The blue line represents the average per-flow delay prediction performance of 10 retraining realizations (in terms of NMSE) with the proposed synchronization elements. The gray dashed line shows the performance without them, and the light blue shaded area indicates the standard deviation of NMSE across all retraining realizations. Finally, the red-shaded area represents the region in which the retraining process occurred across realizations.}
    \label{fig:exp_5g_crosshaul_germany}
\end{figure}

In the first experiment, we observed that concept drift was successfully detected in all instances of deviation caused by changes in the traffic pattern. However, varying detection delays were identified between the moment when the drift occurred and the moment it was detected. The resulting performance is presented in Figure~\ref{fig:exp_5g_crosshaul_germany}, which shows the \ac{VTwin} performance over time in the 5G-Crosshaul and Germany topology in terms of \ac{NMSE}. The figure compares our proposed synchronization elements against a baseline \ac{VTwin} without them, where the baseline scenario can be interpreted as one in which every concept drift results in a false negative.

Before the first concept drift is detected, the \ac{VTwin} operates in a typical manner, as seen in the related work: a \ac{GNN} model that is already trained to generalize to unseen datasets and, based on its predictions, performs the required analysis within the proposed application. However, after the first concept drift occurs, the scenario changes due to traffic variability, leading to model degradation, evidenced by some sample windows exhibiting \ac{NMSE} values near $0$ \si{dB}. Despite this decline, the \ac{VTwin} model remains stable, even without retraining, as the underlying data distributions, such as Poisson and Exponential, are still related. This behavior changes after the second concept drift, as the input data distribution changes to a packet size with a distinct pattern, requiring retraining, which is evident in the accentuated \ac{NMSE} difference between the models with and without retraining. In the last concept drift, the \ac{VTwin} becomes obsolete without a retraining process, as the input data distribution shifts to a \ac{UDP} connection characterized by a constant packet size and the absence of congestion control. At this point, retraining becomes extremely necessary, as demonstrated by the pronounced performance gap between the models with and without retraining.

\begin{figure}[!h]
    \centering
    \includegraphics[scale=0.45]{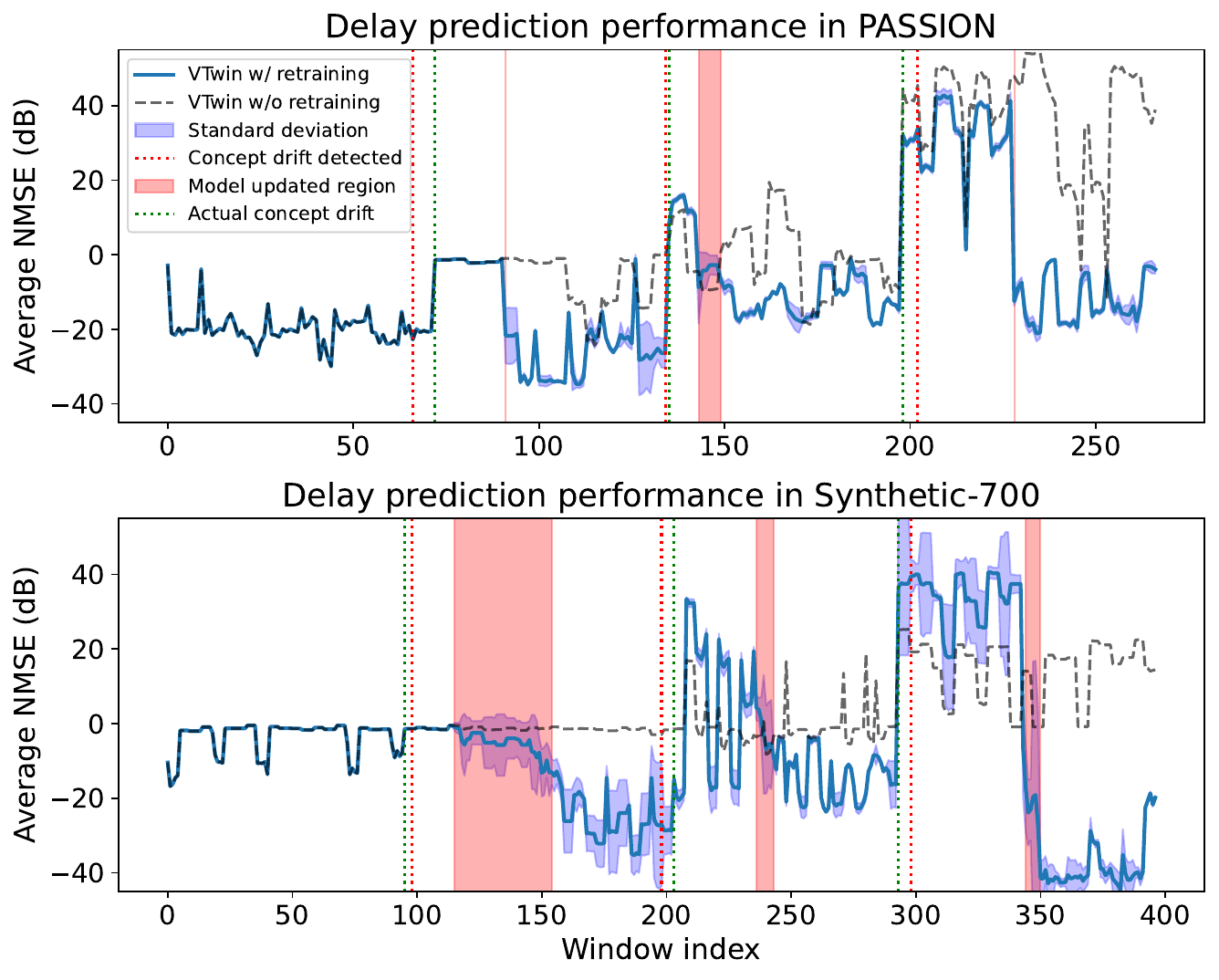}
    \caption{Comparison of the per-flow delay prediction performance using the proposed architecture considering PASSION and Synthetic-700 topologies.}
    \label{fig:experiment_ger_passion}
\end{figure}

The results obtained for the remaining two topologies are intended to further validate the effectiveness of the proposed synchronization method across networks with diverse characteristics in terms of scalability. Specifically, the PASSION topology comprises 128 nodes and approximately three times more links than the 5G-Crosshaul topology, while the Synthetic-700 topology contains roughly thirteen times more nodes than the Germany topology. Despite these substantial differences in scale and connectivity, Figure \ref{fig:experiment_ger_passion} shows consistent behavior across both topologies, comparable to that observed in the 5G-Crosshaul and Germany scenarios. In particular, during the final concept drift detected by the drift detector, the results reinforce the same conclusions regarding the necessity of mandatory retraining due to degraded \ac{VTwin} prediction performance.

\begin{figure}[!h]
    \centering
    \includegraphics[scale=0.6]{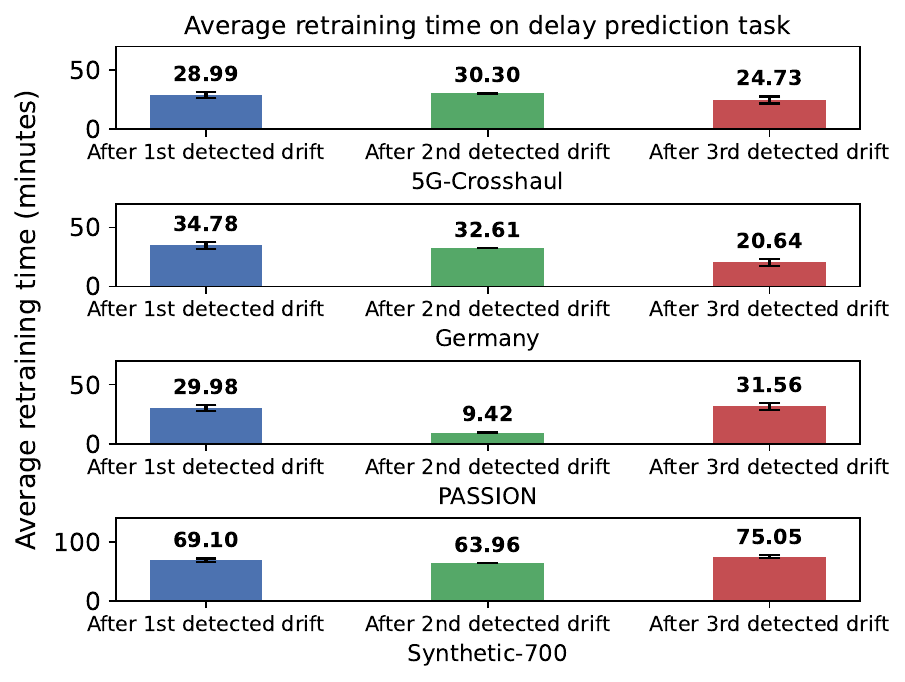}
    \caption{Average retraining time (in minutes) cost by the \ac{VTwin} to synchronize with the \ac{PTwin} in the delay prediction task. The value above each bar represents the average retraining time.}
    \label{fig:VTwin_retrainig_time}
\end{figure}

In terms of retraining time cost in minutes, across these 10 realizations, we obtained the following results depicted by Figure~\ref{fig:VTwin_retrainig_time}. In our environment, the training time analysis shows that three out of four retraining processes for the delay prediction task are completed in less than $35$ minutes. The exception is the Synthetic-700 topology, which is explained by the size of the topology. Considering the overall \ac{NDT} operation time, the feasibility of the retraining duration depends on the specific application, and optimizing this time lies beyond the scope of this work. In our scenario, for instance, within the 5G-Crosshaul topology, 83.41 minutes are spent in retraining states out of a total of 418.8 minutes, corresponding to approximately 20\% of the total operation time. For the Germany and PASSION topologies, the proportion of time devoted to retraining remains below 22\%. Finally, for the Synthetic-700, the retraining time spent was below 32\%, indicating a reasonable balance between adaptability and operational continuity.

Noteworthy, in certain concept drift events, the retraining time can be considerably shorter. For instance, in the PASSION topology, following the second detected drift, the average retraining duration was only $9.42$ minutes. This reduction is primarily due to the use of an \textit{early stopping} callback, which prevents unnecessary training of the \ac{VTwin} model once convergence is reached.

Finally, Table \ref{tab:main_results} summarizes the \ac{NMSE} across all window examples, highlighting \ac{VTwin}'s performance before and after each concept drift. For the \ac{NMSE} after the concept drift, we assume the mean average \ac{NMSE} for comparison with the approach without synchronization, since our evaluation considers 10 retraining process realizations. Hence, incorporating our \ac{NDT} synchronization module results in a performance improvement of at least $64\%$ in predictions compared to configurations without it. Furthermore, given our empirical goal of achieving an average \ac{NMSE} below $-10$~\si{dB}, the approach without \ac{NDT} synchronization meets this target in only 2 of the 12 concept drifts across all evaluated topologies. In contrast, our proposed approach with \ac{NDT} synchronization achieves this performance in 6 out of 12 cases following concept drift. Thus, from a general perspective, for all topologies and after all concept drifts, the performance of the predictions has improved, highlighting the importance of twin synchronization in production environments where concept drift is pervasive.

\begin{table*}[!h]
\centering
\caption{\ac{NMSE} performance in predicting per-flow delay (in \si{dB}) before and after concept drift, considering the \acs{NDT} operation both with and without the proposed synchronization module.}
\label{tab:main_results}

\scalebox{0.9}{\begin{tabular}{lccccc}
\toprule
Topology & Drift ID  & \makecell{Pre-drift avg.\\NMSE w/o sync.} & \makecell{Post-drift avg. \\ NMSE w/o sync.} & \makecell{Pre-drift mean avg. \\ NMSE w/ sync.} & \makecell{Post-drift mean avg.\\NMSE w/ sync.} \\
\midrule
 \multirow{4}{*}{5G-Crosshaul} & 1 & $-20.69$ & $-14.36$ & $-20.69$ & $\mathbf{-18.87}$\\ \cmidrule{2-6}
& 2 & $-14.36$ & $-5.11$ & $-18.87$ & $\mathbf{-10.10}$\\ \cmidrule{2-6}
& 3 & $-5.11$ & $17.40$ & $-10.10$ & $\mathbf{2.08}$ \\
 \midrule
 \multirow{4}{*}{Germany} & 1 & $-18.88$ & $-14.10$ & $-18.88$ & $\mathbf{-19.01}$\\ \cmidrule{2-6}
& 2 & $-14.10$ & $-1.79$ & $-19.01$ & $\mathbf{-13.82}$\\ \cmidrule{2-6}
& 3 & $-1.79$ & $24.64$ & $-13.82$ & $\mathbf{-3.27}$ \\
 \midrule
 \multirow{4}{*}{PASSION} & 1 & $-19.95$ & $-5.41$  & $-19.95$ & $\mathbf{-19.01}$\\\cmidrule{2-6}
& 2 & $-5.41$ & $-1.14$ & $-19.01$ & $\mathbf{-8.06}$\\ \cmidrule{2-6}
& 3 & $-1.14$ & $38.09$ & $-8.06$ & $\mathbf{7.19}$ \\
 \midrule
 \multirow{4}{*}{Synthetic-700} & 1 & $-3.76$ & $-1.44$  & $-3.76$ & $\mathbf{-13.77}$\\\cmidrule{2-6}
& 2 & $-1.44$ & $-0.93$ & $-13.77$ & $\mathbf{-6.42}$\\ \cmidrule{2-6}
& 3 & $-0.93$ & $14.89$ & $-6.42$ & $\mathbf{-2.05}$ \\
 \bottomrule
\end{tabular}}
\end{table*}

\subsection*{Jitter prediction performance}
\begin{figure}[!h]
    \centering
    \includegraphics[scale=0.45]{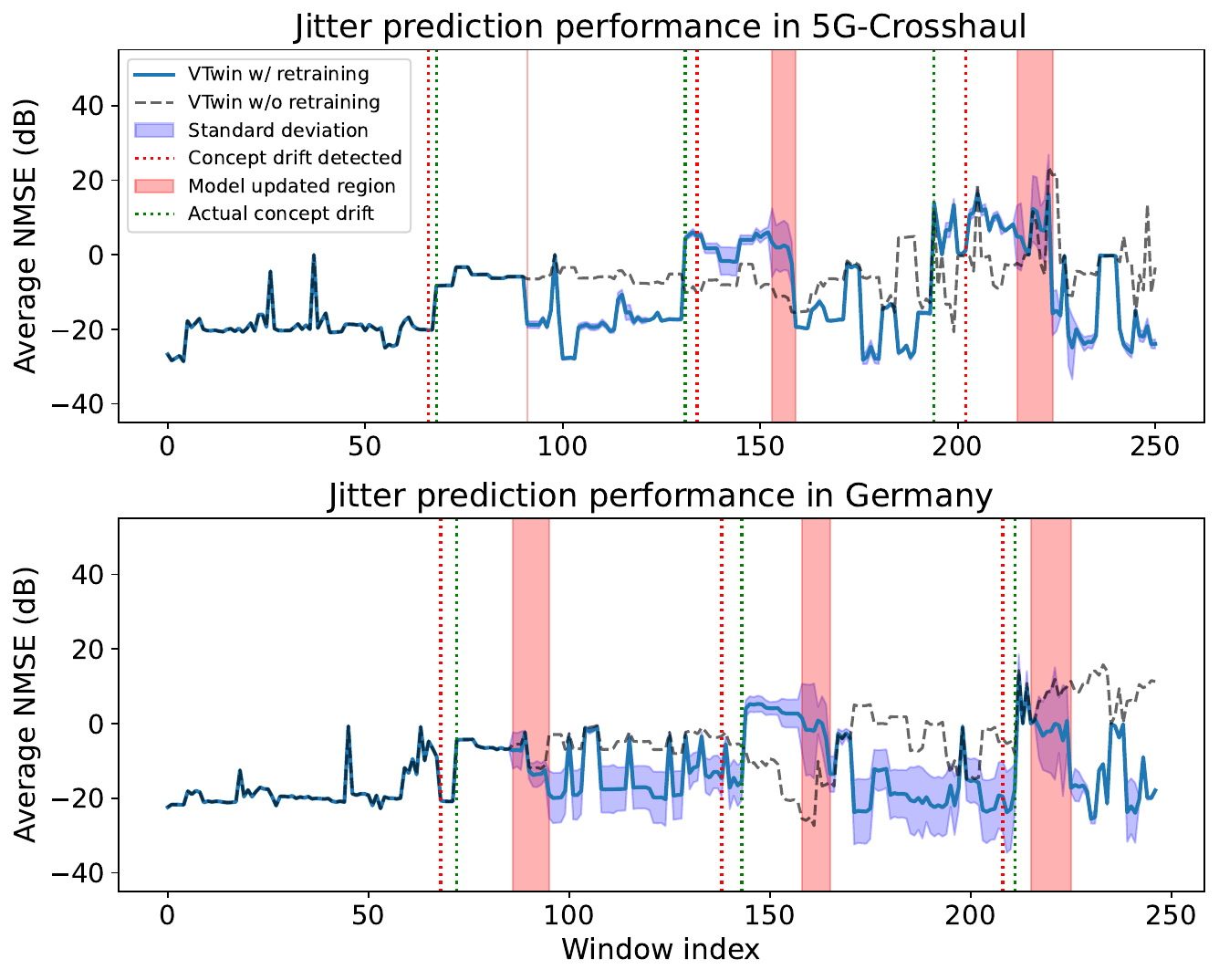}
    \caption{Comparison of the jitter prediction performance using the proposed architecture considering 5G-Crosshaul and Germany topologies.}
    \label{fig:exp_5g_crosshaul_germany_jitter}
\end{figure}

\begin{figure}[!h]
    \centering
    \includegraphics[scale=0.45]{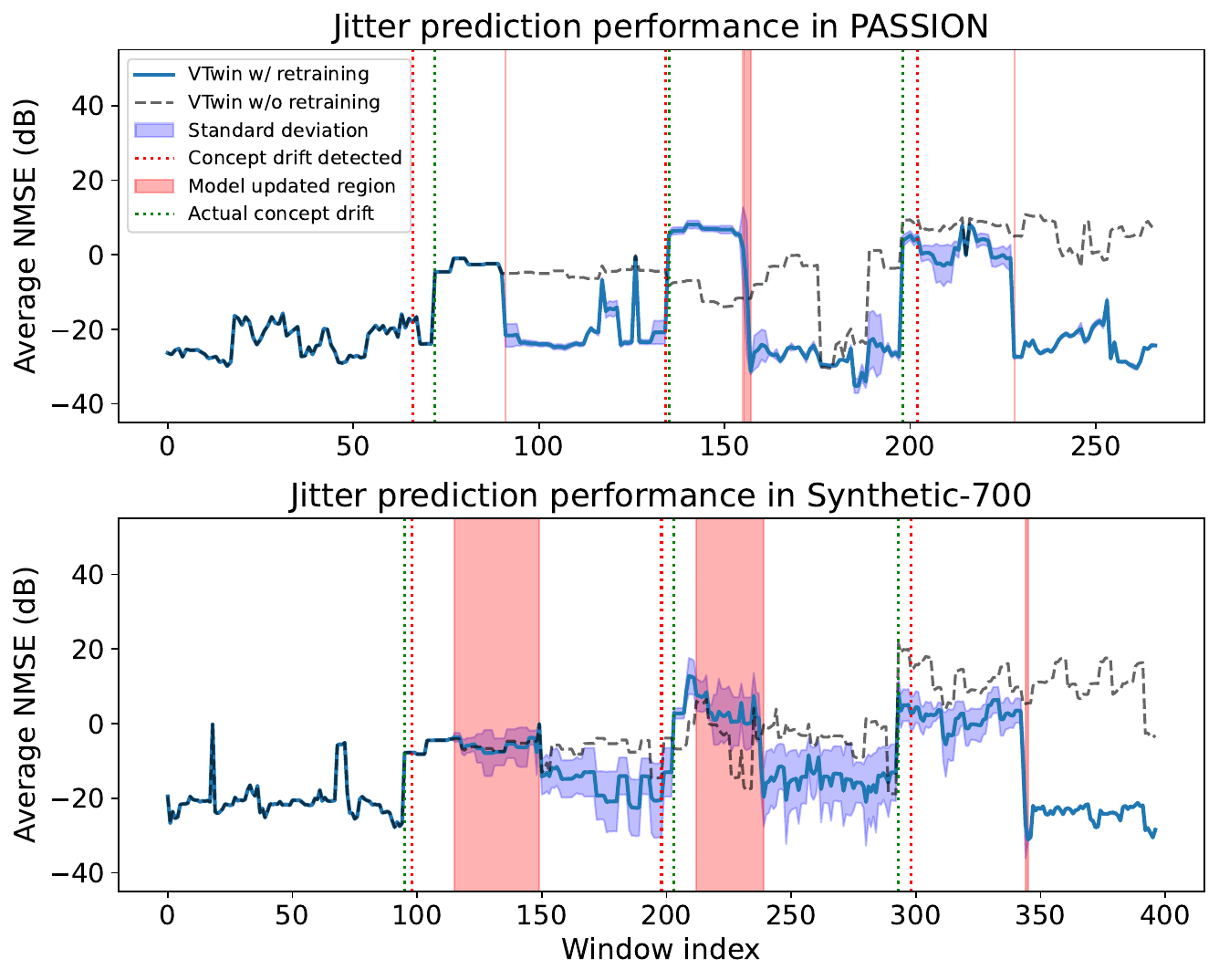}
    \caption{Comparison of the jitter prediction performance using the proposed architecture considering PASSION and Synthetic-700 topologies.}
    \label{fig:experiment_ger_passion_jitter}
\end{figure}

Using the same environment, we also evaluated the per-flow jitter prediction performance in scenarios with and without \ac{NDT} synchronization. In this sense, Figures~\ref{fig:exp_5g_crosshaul_germany_jitter} and~\ref{fig:experiment_ger_passion_jitter} depict the performance in terms of \ac{NMSE} in the 5G-Crosshaul, Germany, PASSION, and Synthetic-700 topologies. From these results, it is possible to perceive similar trends with \ac{NDT} synchronization. For example, after the third detected concept drift, the retraining process is extremely necessary.

\begin{figure}[!h]
    \centering
    \includegraphics[scale=0.6]{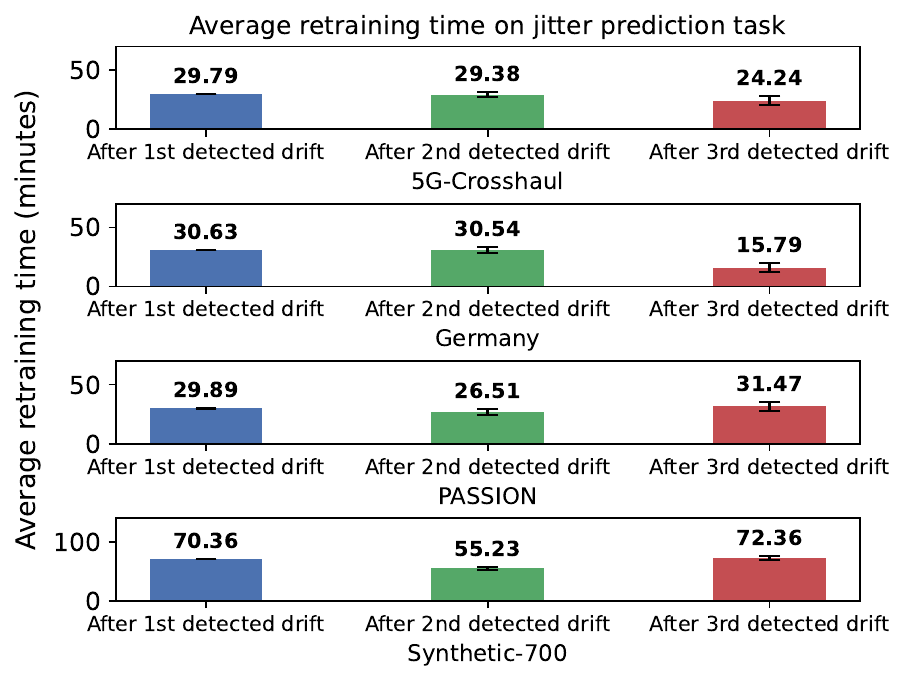}
    \caption{Average retraining time (in minutes) cost by the \ac{VTwin} to synchronize with the \ac{PTwin} in the jitter prediction task. The value above each bar represents the average retraining time.}
    \label{fig:VTwin_retrainig_time_jitter}
\end{figure}

Furthermore, a similar pattern to that observed in the delay prediction training cost can be seen in the average retraining time after a concept drift event for the jitter prediction task, as shown in Figure~\ref{fig:VTwin_retrainig_time_jitter}. In this case, with the exception of experiments with the Synthetic-700 topology, all retraining times remained below $35$ minutes, resulting in a retraining time proportion below $20\%$ for 5G-Crosshaul, Germany, PASSION, and below $30\%$ for the Synthetic-700 topology.

These results for the jitter prediction task reaffirm the important role of the proposed \ac{NDT} synchronization architecture in maintaining satisfactory performance on \ac{QoS} prediction. As shown in Table~\ref{tab:main_results_jitter}, when using our synchronization approach, we achieved an \ac{NMSE} below $-10~\si{dB}$ in 8 out of 12 drift events, whereas the scenario without synchronization reached this level in only one event.

\begin{table*}[!h]
\centering
\caption{\ac{NMSE} performance in predicting per-flow jitter (in \si{dB}) before and after concept drift, considering the \acs{NDT} operation both with and without the proposed synchronization module.}
\label{tab:main_results_jitter}

\scalebox{0.9}{\begin{tabular}{lccccc}
\toprule
Topology & Drift ID  & \makecell{Pre-drift avg.\\NMSE w/o sync.} & \makecell{Post-drift avg. \\ NMSE w/o sync.} & \makecell{Pre-drift mean avg. \\ NMSE w/ sync.} & \makecell{Post-drift mean avg.\\NMSE w/ sync.} \\
\midrule
 \multirow{4}{*}{5G-Crosshaul} & 1 & $-19.78$ & $-5.84$ & $-19.78$ & $\mathbf{-13.66}$\\ \cmidrule{2-6}
& 2 & $-5.84$ & $-7.60$ & $-13.66$ & $\mathbf{-8.67}$\\ \cmidrule{2-6}
& 3 & $-7.60$ & $-1.44$ & $-8.67$ & $\mathbf{-3.83}$ \\
 \midrule
 \multirow{4}{*}{Germany} & 1 & $-18.30$ & $-5.23$ & $-18.30$ & $\mathbf{-12.01}$\\ \cmidrule{2-6}
& 2 & $-5.23$ & $-7.63$ & $-12.01$ & $\mathbf{-11.43}$\\ \cmidrule{2-6}
& 3 & $-7.63$ & $7.88$ & $-11.43$ & $\mathbf{-9.46}$ \\
 \midrule
 \multirow{4}{*}{PASSION} & 1 & $-23.10$ & $-4.17$  & $-23.10$ & $\mathbf{-15.78}$\\\cmidrule{2-6}
& 2 & $-4.17$ & $-10.32$ & $-15.78$ & $\mathbf{-15.68}$\\ \cmidrule{2-6}
& 3 & $-10.32$ & $6.37$ & $-15.68$ & $\mathbf{-12.98}$ \\
 \midrule
 \multirow{4}{*}{Synthetic-700} & 1 & $-20.89$ & $-6.39$  &$-20.89$ & $\mathbf{-11.02}$\\\cmidrule{2-6}
& 2 & $-6.39$ & $-4.30$ & $-11.02$ & $\mathbf{-7.55}$\\ \cmidrule{2-6}
& 3 & $-4.30$ &$11.23$ & $-7.55$ & $\mathbf{-11.51}$ \\
 \bottomrule
\end{tabular}}
\end{table*}

\subsubsection*{Use case with SLA violation monitoring}
The use case scenario employed to evaluate the proposed \ac{NDT} synchronization, considering a high-level metric related to \ac{SLA} monitoring tasks, was conducted using data derived from the four topologies used in the last experiments. In this sense, we evaluated the same scenario used to validate the performance on prediction per-flow delay, but we considered high-level metrics such as the accuracy of the \ac{SLA} violation predictions. In this case, we assume a simplified \ac{SLA} compliance scenario, admitting only the predicted per-flow delay as the metric used to make a decision regarding the \ac{SLA} violation. Thus, we assume that each flow has a specific \ac{PDB} value that was assigned empirically based on the characteristics of each topology, such as the propagation delay of each link.

Regarding the experiments, we considered a total of $25\,100$, $24\,600$, $26\,700$, and $39\,700$ flows traversing the 5G-Crosshaul, Germany, PASSSION, and Synthetic-700 network topologies, respectively. Each flow was evaluated by the \ac{SLA} monitoring application to determine whether it was compliant with or in violation of the \ac{PDB}, based on the delay predicted by the \ac{VTwin}. An \ac{SLA} violation occurs when the per-flow delay exceeds the \ac{PDB} for the corresponding path. Table~\ref{tab:topologies_sla_char} summarizes the number of flows identified as being in violation for each topology.

\begin{table}[!h]
    \centering
    \caption{Flow traffic characteristics for each topology used for \ac{SLA} compliance classification.}
    
    \scalebox{1}{\begin{tabular}{lcc}
    \toprule
      Topology   & Total number of flows & Number of flows in violation  \\
    \midrule
       5G-Crosshaul & $25\,100$ & $536$ \\ 
       Germany & $24\,600$ & $793$ \\ 
       PASSION & $26\,700$ & $1\,229$ \\ 
       Synthetic-700 & $39\,700$ & $1\,527$ \\ 
    \bottomrule
    \end{tabular}}
    \label{tab:topologies_sla_char}
\end{table}

\begin{figure}[!h]
    \begin{subfigure}[b]{0.48\textwidth}
    \centering
    \caption{Accuracy of SLA violation predictions in 5G-Crosshaul topology.}
    \includegraphics[scale=0.38]{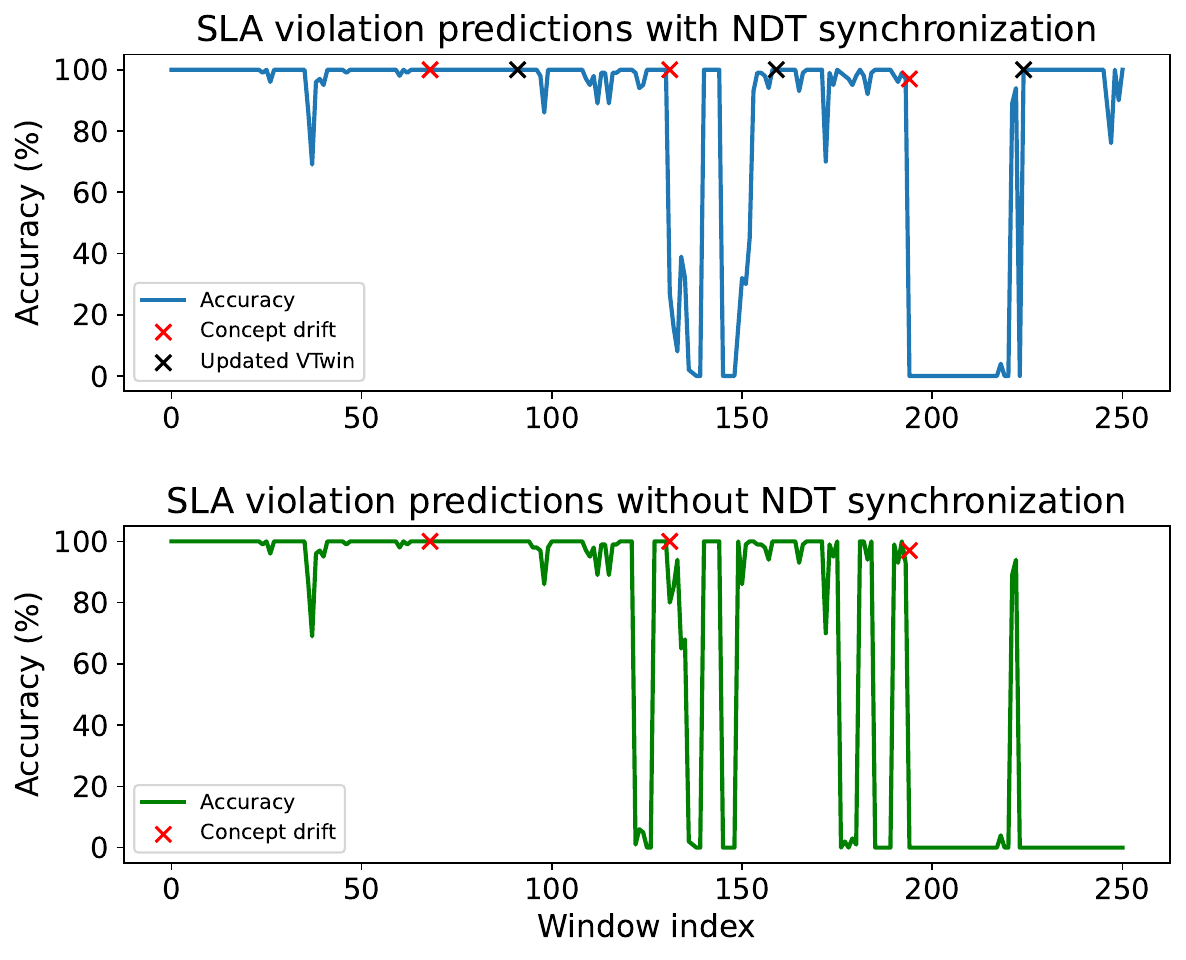}
    \label{fig:sla_violation_5g_crosshaul}
    \end{subfigure}    
    \begin{subfigure}[b]{0.5\textwidth}
    \centering
    \caption{Accuracy of SLA violation predictions in Germany topology.}
    \includegraphics[scale=0.38]{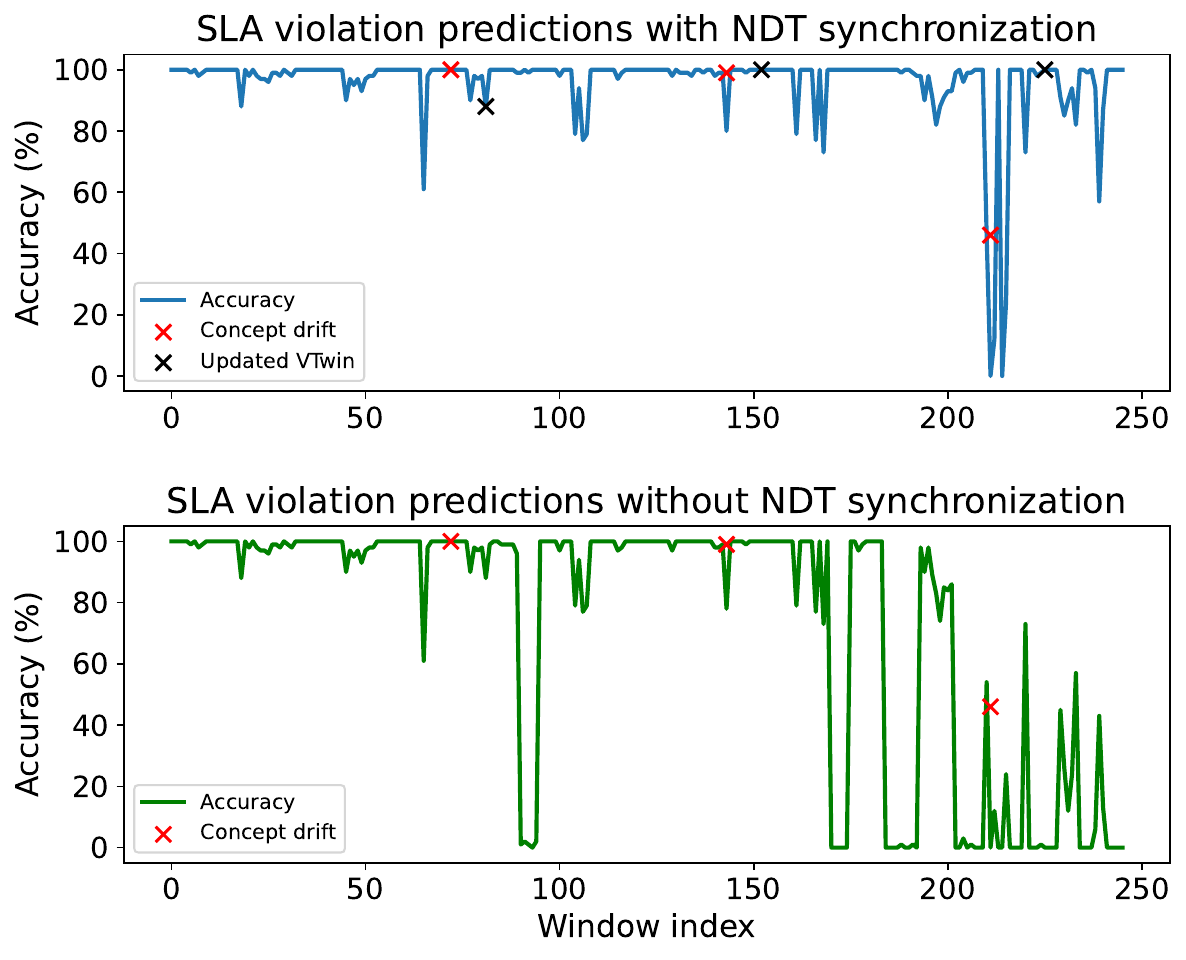}
    \label{fig:sla_violation_germany}
    \end{subfigure}
\caption{Accuracy of \ac{SLA} violation classification considering 5G-Crosshaul and Germany topologies.}
\end{figure}

The accuracy in the prediction of \ac{SLA} violations is shown in Figure~\ref{fig:sla_violation_5g_crosshaul} and~\ref{fig:sla_violation_germany}, which considers the \ac{VTwin} with and without the \ac{NDT} synchronization module in the 5G-Crosshaul and Germany topologies. Each window index corresponds to the classification of 100 flows for better visualization of the results in this figure. In all topologies, the application exhibits the expected degradation in classification performance when the \ac{NDT} synchronization module is disabled. This occurs because the \ac{VTwin} states are not updated with those of the \ac{PTwin}. Quantitatively, this results in $8\,115$ and $6\,347$ flows being incorrectly classified as compliant with their respective \acp{SLA}, in the 5G-Crosshaul and Germany topologies, respectively. 

\begin{figure}[!h]
    \begin{subfigure}[b]{0.48\textwidth}
    \centering
    \caption{Accuracy of \ac{SLA} violation predictions in PASSION topology.}
    \includegraphics[scale=0.38]{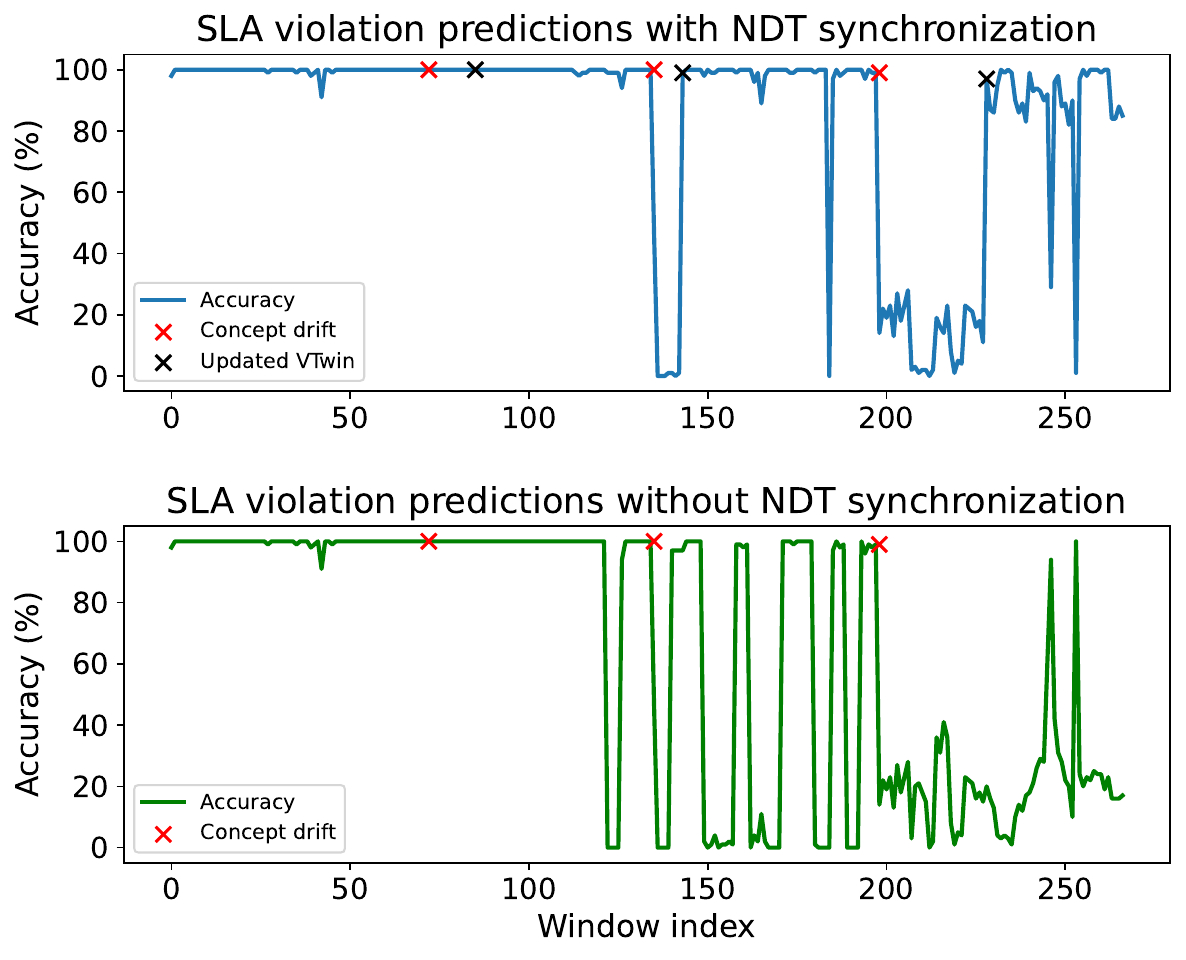}
    \label{fig:sla_violation_passion}
    \end{subfigure}    
    \begin{subfigure}[b]{0.5\textwidth}
    \centering
    \caption{Accuracy of \ac{SLA} violation predictions in Synthetic-700 topology.}
    \includegraphics[scale=0.38]{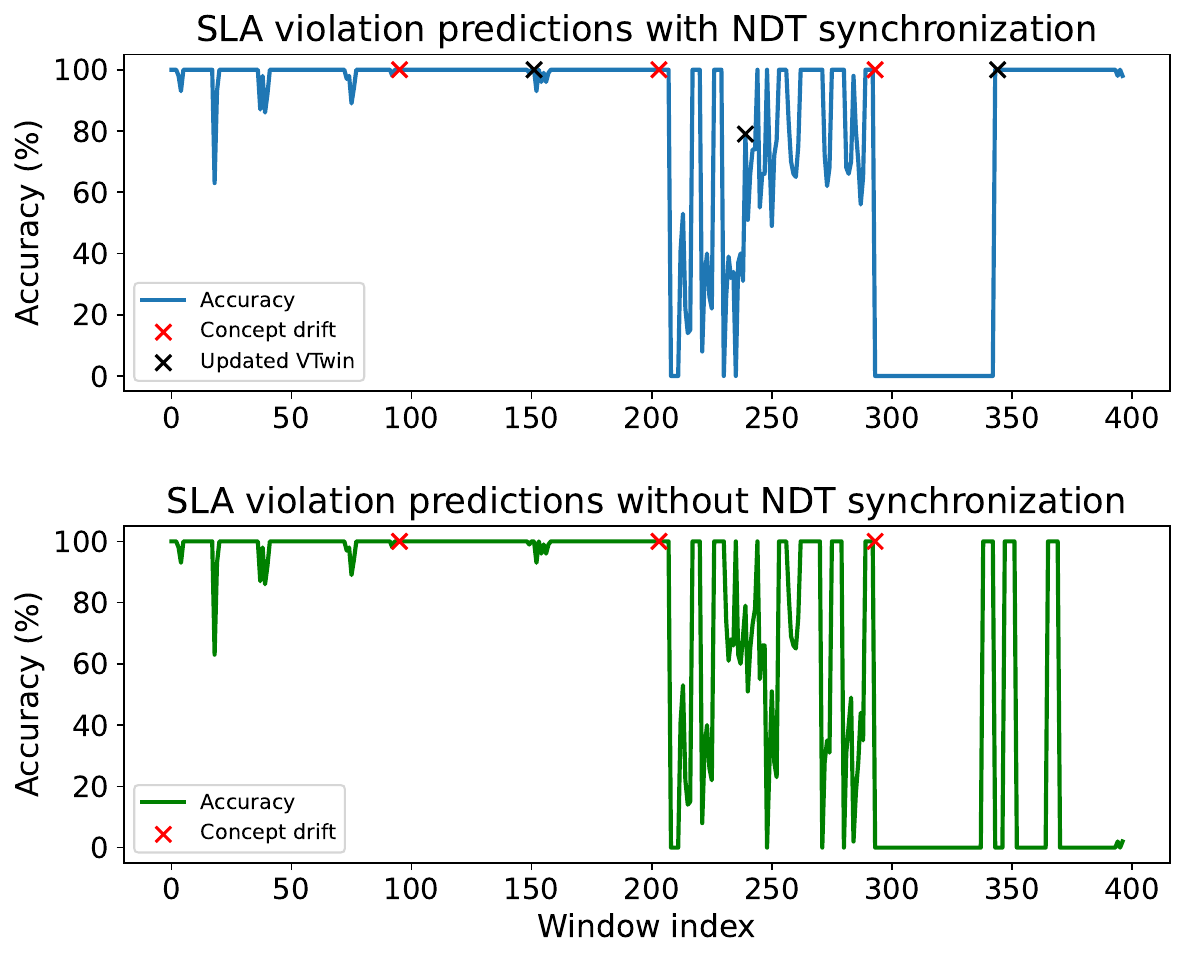}
    \label{fig:sla_violation_random}
    \end{subfigure}
\caption{Accuracy of \ac{SLA} violation classification considering PASSION and Synthetic-700 topologies.}
\end{figure}

When the \ac{NDT} synchronization module is enabled, however, the number of misclassified flows decreases significantly to $4\,536$ and $982$, respectively. Similar improvements are observed in the other two topologies, as depicted in Figure~\ref{fig:sla_violation_passion} and \ref{fig:sla_violation_random}, respectively, the PASSION and Synthetic-700 topologies. Moreover, all numerical results are summarized in
Table~\ref{tab:sla_violations_results}. It is also noteworthy that part of these \ac{SLA} violations occurs during periods in which the \ac{VTwin} is undergoing retraining. Overall, this experiment demonstrates that the practical usability of an \ac{NDT} critically depends on an effective synchronization mechanism between the twins operating at an appropriate rate, which is essential for transitioning the solution from a prototype to a production-ready framework.

\begin{table}[!h]
    \centering
    \caption{Application performance on \ac{SLA} violation classification with and without \ac{NDT} synchronization.}
    
    \scalebox{0.9}{\begin{tabular}{lccc|ccc}
    \toprule
    & \multicolumn{3}{c|}{Without NDT synchronization} & \multicolumn{3}{c}{With NDT synchronization}\\ 
    \midrule
      Topology & \makecell{Number of flows\\ misclassified} &  \makecell{Number of flows\\ correctly classified} & \makecell{Accuracy\\(\%)} & \makecell{Number of flows\\ misclassified} &  \makecell{Number of flows\\ correctly classified} & \makecell{Accuracy\\(\%)} \\
    \midrule
       5G-Crosshaul & $8\,115$ & $16\,985$ & $67.66$ & $\mathbf{4\,536}$ & $\mathbf{20\,564}$ &  $\mathbf{81.92}$ \\ 
       Germany & $6\,347$ & $18\,253$ & $74.19$ & $\mathbf{982}$ & $\mathbf{23\,618}$ & $\mathbf{96.00}$ \\ 
       PASSION & $9\,019$ & $17\,681$ & $66.22$ & $\mathbf{3\,941}$ & $\mathbf{22\,759}$ & $\mathbf{85.23}$ \\ 
      Synthetic-700 & $12\,133$ & $27\,597$ & $69.43$ & $\mathbf{7\,788}$ & $\mathbf{31\,912}$ & $\mathbf{80.38}$ \\ 
    \bottomrule
    \end{tabular}}
    \label{tab:sla_violations_results}
\end{table}

\section{Conclusions and Future Works}
\label{sec:conclusion}

This article proposed an enhanced \ac{NDT} architecture incorporating data-driven synchronization modules, taking initial steps toward addressing practical and production-level aspects of \ac{NDT} deployment in transport networks. The proposed modules within this architecture specifically tackle the challenge posed by the non-stationary nature of network traffic over time. In this sense, this architecture aims to ensure synchronization between the two twins of the \ac{NDT} platform. To achieve this, the \ac{NDT} synchronization modules, such as the concept drift detector, aim to track and detect possible changes in the traffic pattern, triggering a retraining process and ultimately updating the \ac{VTwin} weights through the \ac{VTwin} model management. To evaluate the architecture's operation, we considered four different topologies with sizes ranging from 50 to 700 nodes in an environment with flows using four packet size distributions, generating various traffic patterns. Thus, to detect and initiate the retraining process, we considered an input distribution-based approach, which was based on the \ac{KSWIN} method. Our results demonstrate the importance of the synchronization process between the \ac{VTwin} and \ac{PTwin}, achieving a performance improvement in predicting total per-flow and per-flow jitter of at least $64\%$ and $21\%$, respectively, compared to a configuration without a concept drift detector. Finally, aimed at obtaining insight into the impact of the proposed \ac{NDT} architecture at a high level, we also evaluated the \ac{NDT} synchronization, considering an application related to \ac{SLA} monitoring.

The results of the proposed experiments offer valuable insights into the challenges of twin synchronization, positioning this article as a foundational step toward future developments. A key direction for future work involves progressively incorporating more realistic elements into the experimental environments. One major limitation is the gap between the current setup and real-world production environments. Although the proposed module presents a promising approach to addressing specific challenges faced by \ac{NDT} systems, it does not fully resolve all the issues required for the complete feasibility of the \ac{NDT} platform. For instance, while the data management module includes essential mechanisms for collecting information from the \ac{PTwin}, the experiments rely on traffic generated by a traffic generator, with metrics directly obtained from simulator logs. In contrast, data collection in a production environment is significantly more complex and may require integration with auxiliary tools, such as those defined by protocols such as the Simple Network Management Protocol (SNMP), which is used to collect network metrics. Furthermore, the feasibility of the proposed synchronization method also depends on the ability to generate synthetic data to retrain the \ac{VTwin} model when necessary, which is not addressed in this article. In this context, the use of generative \ac{AI} emerges as a promising solution and represents a potential avenue for future research.

In addition to addressing the data management tasks inherent to an \ac{NDT}, future research will also focus on enhancing the synchronization method itself, which can be explored along two main directions. In the short term, we aim to assess synchronization performance under varying transport network conditions by adopting a more robust multivariate concept drift detection approach, one that simultaneously accounts for multiple input features, such as traffic rate, per-link load, and link capacity. Moreover, in future work, we will also consider incorporating a more realistic distribution of packet rates, as the current study assumes the same distribution for packet rate and packet size. Another aspect that should be evaluated in future work is the performance impact of the \ac{VTwin} retraining process caused by false positives from these concept drift detectors, which is not addressed in this present work. In the long term, we aim to develop a more proactive synchronization strategy by integrating online learning techniques. This approach would reduce the latency in adapting the \ac{VTwin} model parameters by eliminating the need to wait for explicit concept drift detection, thereby minimizing the period during which outdated model weights remain in use.

\section*{CRediT authorship contribution statement}
\textbf{Cláudio Modesto}: Writing --- original draft, Writing --- review and editing, Visualization, Validation, Software, Methodology, Investigation, Data curation, Conceptualization. \textbf{João Borges}: Writing --- original draft, Conceptualization. \textbf{Cleverson Nahum}: Software, Supervision, Writing – review and editing. \textbf{Lucas Matni}: Software, Formal analysis. \textbf{Cristiano Bonato Both}: Writing – review and editing, Supervision, Visualization. \textbf{Kleber Cardoso}: Writing – review and editing, Methodology. \textbf{Glauco Gonçalves}: Writing – review and editing, Formal analysis. \textbf{Ilan Correa}: Writing – review and editing. \textbf{Silvia Lins}: Writing – review and editing, Project administration. \textbf{Andrey Silva}: Writing – review and editing, Project administration. \textbf{Aldebaro Klautau}: Writing – review and editing, Project administration, Funding acquisition.

\section*{Acknowledgments}
This study was financed in part by Coordenação de Aperfeiçoamento de Pessoal de Nível Superior - Brasil (CAPES) – Finance Code  001, the Conselho Nacional de Desenvolvimento Científico e Tecnológico (CNPq); the Innovation Center, Ericsson Telecomunicações Ltda., Brazil; and Project Smart 5G Core and MUltiRAn Integration (SAMURAI) (MC-TIC/CGI.br/FAPESP under grant 2020/05127-2).

%% If you have bib database file and want bibtex to generate the
%% bibitems, please use
%%
%%  \bibliographystyle{elsarticle-num} 
%%  \bibliography{<your bibdatabase>}

%% Refer following link for more details about bibliography and citations.
%% https://en.wikibooks.org/wiki/LaTeX/Bibliography_Management

%\textcolor{blue}{Referência com mais de 3 autores, precisam estar com PrimeiroAutor et al.}

%%`Elsevier LaTeX' style
\bibliographystyle{elsarticle-num}
\bibliography{references}

\end{document}